\documentclass[preprint2,longabstract]{aastex}  
\usepackage{verbatim}
\usepackage{natbib}
\usepackage{amssymb}
\usepackage{times}
\usepackage{pstricks}
\usepackage{epsfig}

\def\ls{\mathrel{\raise0.27ex\hbox{$<$}\kern-0.70em 
\lower0.71ex\hbox{{$\scriptstyle \sim$}}}}
\def\gs{\mathrel{\raise0.27ex\hbox{$>$}\kern-0.70em 
\lower0.71ex\hbox{{$\scriptstyle \sim$}}}}

\newcommand{\um}{10}
\newcommand{\cit}{15}
\newcommand{\lbl}{1}
\newcommand{\stockholm}{3}
\newcommand{\lpnhe}{5}
\newcommand{\yale}{6}
\newcommand{\lam}{7}
\newcommand{\upenn}{8}
\newcommand{\ucb}{9}
\newcommand{\stsci}{11}
\newcommand{\cppm}{12}
\newcommand{\iu}{13}
\newcommand{\aas}{14}
\newcommand{\cwr}{16}
\newcommand{\cambridge}{17}
\newcommand{\cea}{18}
\newcommand{\ipnl}{19}
\newcommand{\slac}{2}
\newcommand{\fnal}{4}

\begin{document}

\title{Supernova /  Acceleration Probe: A Satellite Experiment
to Study the Nature of the Dark Energy} 

\renewcommand{\thefootnote}{\fnsymbol{footnote}}
\author{
G.Aldering\altaffilmark{\lbl}, W.~Althouse\altaffilmark{\slac},
R.~Amanullah\altaffilmark{\stockholm}, J.~Annis\altaffilmark{\fnal},
P. Astier\altaffilmark{\lpnhe}, C.~Baltay\altaffilmark{\yale}, E.~Barrelet\altaffilmark{\lpnhe},
S.~Basa\altaffilmark{\lam},
C.~Bebek\altaffilmark{\lbl},
L.~Bergstr\"{o}m\altaffilmark{\stockholm}
% ...
}

\altaffiltext{\lbl}{ Lawrence Berkeley National Laboratory}
\altaffiltext{\slac} {Stanford Linear Accelerator Center}
\altaffiltext{\stockholm}{ University of Stockholm}
\altaffiltext{\fnal}{Fermi National Accelerator Laboratory}
\altaffiltext{\lpnhe}{ LPNHE, CNRS-IN2P3, Paris, France}
\altaffiltext{\yale}{Yale University}
\altaffiltext{\lam}{ LAM, CNRS-INSU, Marseille, France}
\altaffiltext{\upenn}{ University of Pennsylvania}
\altaffiltext{\ucb}{ University of California at Berkeley}
\altaffiltext{\um}{ University of Michigan}
\altaffiltext{\stsci}{ Space Telescope Science Institute}
\altaffiltext{\cppm}{ CPPM, CNRS-IN2P3, Marseille, France}
\altaffiltext{\iu}{ Indiana University}
\altaffiltext{\aas}{ American Astronomical Society}
\altaffiltext{\cit}{ California Institute of Technology}
\altaffiltext{\cwr}{ Case Western Reserve University}
\altaffiltext{\cambridge}{ Cambridge University}
\altaffiltext{\cea}{CEA, Saclay, France}
\altaffiltext{\ipnl}{ IPNL, CNRS-IN2P3, Villeurbanne, France}

\begin{abstract}
The Supernova / Acceleration Probe (SNAP) is a proposed space-based
experiment designed to
study the dark energy and alternative explanations of the acceleration
of the Universe's expansion by 
performing a series of complementary systematics-controlled
astrophysical measurements.
%
%
We here describe a self-consistent reference mission
design that can accomplish this goal
with the two leading measurement approaches being
the Type Ia supernova Hubble diagram and a wide-area
weak gravitational lensing survey.  This design
has been optimized to first order and is now under study
for further modification  and optimization.
A 2-m three-mirror anastigmat wide-field
telescope feeds a focal plane consisting of
a 0.7~square-degree
imager tiled with equal areas of optical CCDs and near infrared
sensors, and
a high-efficiency
low-resolution integral field spectrograph.
The instrumentation suite provides simultaneous discovery
and light-curve measurements of
supernovae and then can
target individual objects for detailed spectral
characterization.  The SNAP mission will discover thousands of
Type~Ia supernovae out to $z=3$ and will obtain high-signal-to-noise
calibrated light-curves and spectra for a subset of $> 2000$
supernovae at redshifts between $z=0.1$ and $1.7$
in a northern field and in a southern field.
A wide-field survey covering one thousand square degrees in both
northern and southern fields resolves
$\sim 100$ galaxies per square arcminute, or a total of more
than 300 million galaxies.
With the PSF stability
afforded by a space observatory, SNAP will provide precise and accurate
measurements of gravitational lensing.
The
high-quality data available in space, combined with the large sample
of supernovae, will enable stringent control of systematic
uncertainties.  
The resulting data
set will be used to determine the energy density of dark energy and 
parameters that describe its dynamical behavior.  The data also
provide a direct test of theoretical models for the dark energy,
including discrimination of
vacuum energy due to the cosmological constant and various classes of
dynamical scalar fields.
If we
assume we
live in a cosmological-constant-dominated
Universe, 
the matter density, dark energy density, and flatness of space can 
all be measured with SNAP supernova and weak-lensing measurements 
to a systematics-limited accuracy of 1\%.  For a flat universe, 
the 
density-to-pressure ratio of dark
energy or equation of state $w(z)$ can be similarly measured to 
5\% for the present value $w_0$ and $\sim0.1$ for the time variation 
$w'\equiv dw/d\ln a|_{z=1}$. 
For a fiducial SUGRA-inspired universe,
$w_0$ and $w'$ can be measured to an even tighter uncertainty of
0.03 and 0.06 respectively.
Note that no external priors are needed. 
As more accurate theoretical
predictions for the small-scale weak-lensing shear develop, the 
conservative estimates adopted here for space-based systematics 
should improve, allowing even tighter
constraints. 
While the survey strategy is tailored for supernova and
weak gravitational lensing observations, the large survey area, depth,
spatial resolution, time-sampling, and nine-band optical to NIR
photometry will support 
additional independent and/or complementary dark-energy measurement
approaches as well as
a broad range of auxiliary science programs.
\end{abstract}
%



\section{Introduction}
\label{sect:intro}  

In the past decade the study of cosmology has taken its first major
steps as a precise empirical science, combining concepts and tools
from astrophysics and particle physics.  The most recent of these
results have already brought surprises.  The Universe's expansion is
apparently accelerating rather than decelerating as expected due to
the gravitational attraction of matter.  This implies that the flat,
matter-dominated model for the Universe does not apply and
that our current fundamental physics understanding of particles,
forces, and fields is likely to be incomplete.

\begin{figure}
\plotone{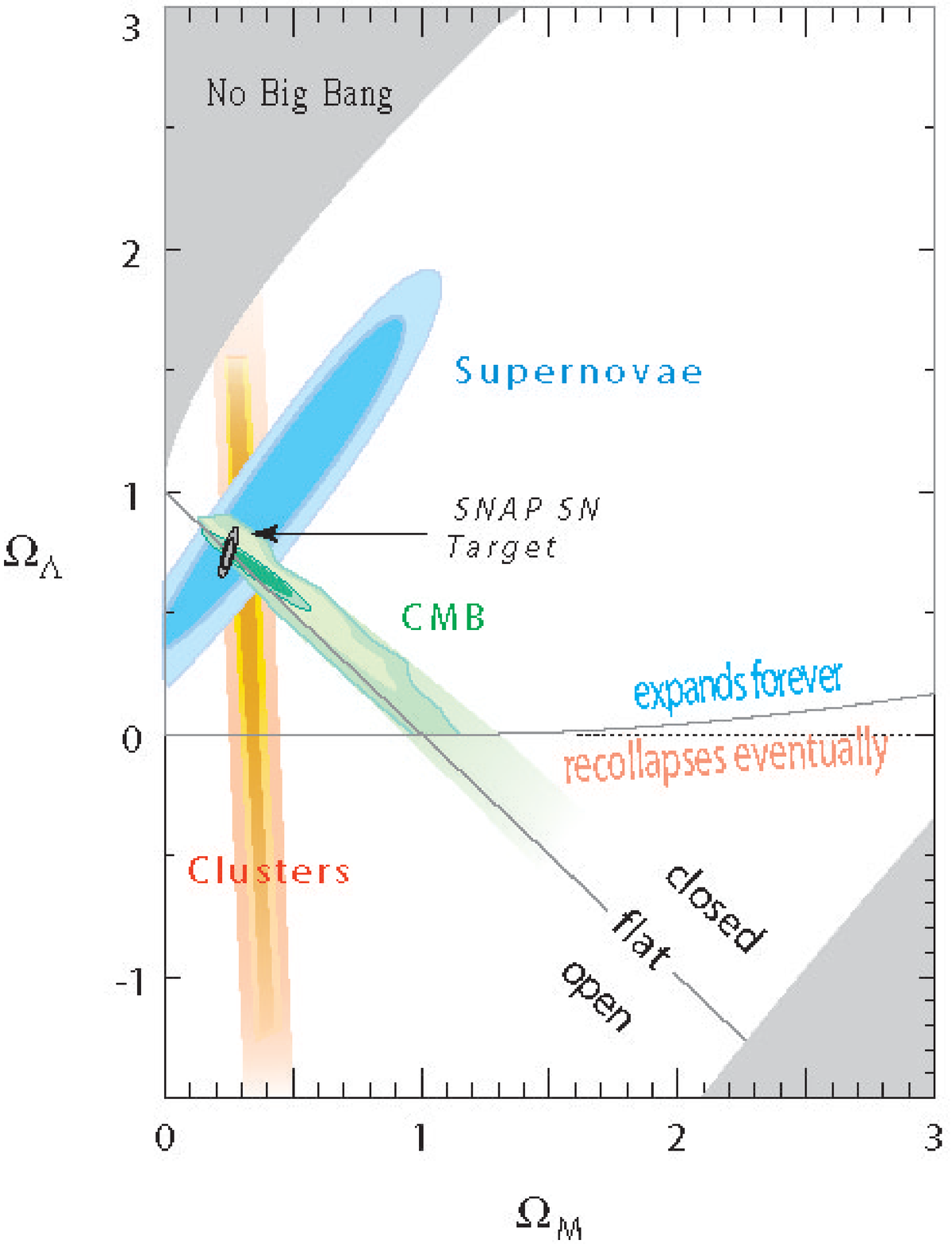}
\caption[$\Omega_M$---$\Omega_\Lambda$ confidence regions with current
supernova, galaxy power spectrum and cluster abundance measurements,
and CMB results.] {There is strong evidence for the
existence of a cosmological vacuum energy density.  Plotted are
$\Omega_M$---$\Omega_\Lambda$ 68\% and 95\%
confidence regions for supernovae
\citep{knopetal:2003}, cluster
measurements (based on \citet{allenetal:2003}), and CMB data
with $H_0$ priors
(outer counters from \citet{Lange:2001},
inner contours from \citet{spergel:2003}).
These results rule
out a simple flat $\Omega_M=1$, $\Omega_\Lambda=0$ cosmology.  Their
consistent overlap is a strong indicator for dark energy.  Also shown
is the expected confidence region from just the SNAP supernova program for
$\Omega_M=0.28, \Omega_\Lambda=0.72$.}
\label{confcmbclust}
\end{figure}

This evidence for a vacuum energy, or more generally a
negative-pressure component called ``dark energy'', comes from the
supernova measurements of changes in the Universe's expansion rate
that directly show the acceleration.  Figure~\ref{confcmbclust} shows
the results of \citet{knopetal:2003} (see also \citet{tonryetal:2003}) which
compare the standardized brightnesses of ground and space-observed
high-redshift  ($0.18<z<0.863$) Type~Ia
supernovae (SNe~Ia) with a large
sample of low-redshift SNe~Ia \citep{hamuyetal:1996,riess_data99}.  The
data implies that for a flat universe a fraction
$\Omega_\Lambda=0.75\pm0.07$ of the
critical density would reside in a cosmological constant.
More generally, these measurements indicate the presence of a
new, unknown energy component with energy density
$\Omega_w$ that can cause acceleration, hence
having an equation of state (ratio of pressure to energy density
\citep{turnerwhite:1997}) with
$w\equiv p/\rho<-1/3$.

The evidence for dark energy has received strong corroboration from 
cosmic microwave background (CMB) results 
\citep{balbi:2000,lange2001,spergel:2003}  which are sensitive to the 
total energy density, combined with galaxy power spectrum and cluster
abundance measurements
\citep{bahcall00,efstathiou:2002,percival:2002,allenetal:2003}
which probe the matter 
density $\Omega_M$,
or with an $H_0$ prior
(see Figure~\ref{confcmbclust}).  Two of these three
independent measurements would have to be in
error to make dark energy unnecessary in the cosmological model.

The dark energy might be due to the cosmological constant term in
Einstein's equations, which implies vacuum energy with
$w=-1$. Alternatively, it could be due to a dynamical scalar field with $w\neq -1$
and/or time-varying $w$.
The
fundamental importance of a universal vacuum energy has sparked a
flurry of activity in theoretical physics with several classes of
models being proposed (e.g.  quintessence 
\citep{ratra_scal_87,cald_Q_98,ferr_scal_98},
Pseudo-Nambu-Goldstone Boson (PNGB) models \citep{frieman95,coble97},
cosmic defects \citep{vilenkin84,vilenkin94},
and modified gravity \citep{DGP:2000,carrolletal:2003}.

\renewcommand{\thefootnote}{\arabic{footnote}}
\setcounter{footnote}{0}

\begin{figure}[ht]
\plotone{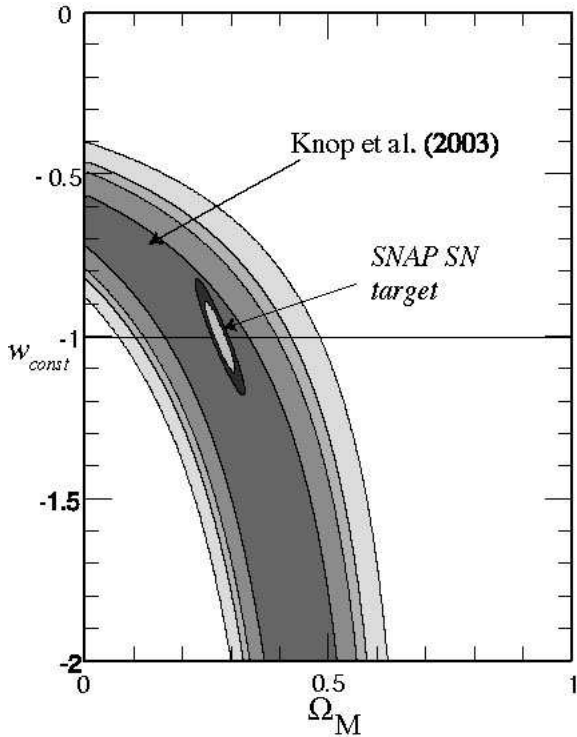}
\caption[Confidence regions in the $\Omega_{\rm M}$--$w$ plane.]
{Best-fit 68\%, 90\%, 95\%, and 99\% confidence regions in the
$\Omega_{\rm M}$--$w$ plane for an additional energy density component,
$\Omega_w$, characterized by an equation-of-state $w = p/\rho$ using
supernovae alone
\citep{knopetal:2003}. The fit is
constrained to a flat cosmology ($\Omega_{\rm M} + \Omega_w =1$).  Also
shown are the expected 68\% and 95\% confidence regions
 allowed by just the SNAP supernova
program assuming
the fiducial values $w=-1$
and $\Omega_M=0.28$.}
\label{wconf}
\end{figure}

All these models explaining the accelerating
Universe have observational consequences for the supernova Hubble diagram
\citep{weller02}.  The luminosity distance as a function of redshift
can for convenience be parameterized to first order
with effective dark-energy parameters, e.g.\ $w_{const}$, or $w_0$, $w_a$.
A constant $w_{const}$ can represent dark-energy models such as the cosmological
constant while
$w=w_0+w_a(1-a)$ is successful at describing a wide variety of both
scalar field and more general models \citep{linder:2003,lindergrav:2004}.
Predictions of dark-energy models can be either compared directly to supernova
Hubble diagrams or to the dark-energy parameters as fit by the
data.
Placing some constraints on possible dark-energy models,
\citet{42SNe_98,garnavich98,ptw99,knopetal:2003} find that for a flat universe, the data
are consistent with a cosmological-constant equation of state and
$0.2\ls\Omega_M\ls 0.4$ (Figure~\ref{wconf}), or generally $w_{const}<-0.6$ at
95\% confidence level.
When combining supernova with other results, such as from a variety of
CMB experiments and large-scale-structure
measurements\citep{hawkins:2002,spergel:2003},
\citet{knopetal:2003} find $w_{const}=-1.05^{+0.15}_{-0.20}$ ($\pm 0.09$
systematics)
and \citet{riessetal:2004} get $w_{const}=-1.02^{+0.13}_{-0.19}$.
Cosmic strings ($w=-1/3$) are already ruled out as dark energy,
while the domain walls ($w=-2/3$) and tracking quintessence
models ($w \gtrsim -0.7$) are disfavored.

Current and upcoming ground-based supernova experiments will continue
to improve measurements of the dynamical effects of the dark energy.
For example, the Nearby Supernova Factory \citep{aldering02} will produce a photometrically calibrated spectral
time-series for hundreds of $0.03<z<0.08$ supernovae, providing a rich
data set for refining our understanding of SN Ia behavior.
The SNLS\footnote{http://snls.in2p3.fr/} and Essence\footnote{http://www.ctio.noao.edu/essence/} projects
will discover and follow hundreds
of supernovae to $\sim 0.03-0.05$ mag accuracy in the range $z =0.3-0.7$,
giving significantly tighter constraints on the value of $w_{const}$.
Unfortunately once the assumption that $w$ is constant is relaxed,
these experiments are not able to provide strong constraints on
$w_0$ nor $w'$
\citep{perlmutterschmidt:2003}.  Testing the nature of dark
energy will thus be difficult with these experiments.

A new experiment is then necessary to study the properties
and possible models for the dark energy, and refine the
measurement of the matter and dark energy densities of the Universe.
Since the discovery of cosmic acceleration in 1998, many
observational approaches to dark energy have been proposed (see
\citet{kujatetal:2002, tegmark:2002, linderiau:2003, cooray:2004}  for recent reviews).
The most well developed and understood method  of constructing
the expansion history of the Universe is the same one that led
to the discovery of dark energy: the measurement of the supernova
Hubble diagram.
The systematic and statistical uncertainties in current SN Ia measurements are
on the same order of magnitude: any new experiment to perform
precision cosmology must eliminate or significantly reduce the influence of the
systematic uncertainties while discovering and measuring
high-quality light curves and spectra of a statistically large number
of supernovae.

In this paper we describe the Supernova/Acceleration Probe (SNAP), a
dedicated satellite mission designed to measure the properties of dark
energy.  Originally conceived in 1999
\citep{perlmutter:1999}, the SNAP concept has progressed
through a series of refinements all starting with science requirements
and then deriving observational, instrumental, and mission
requirements.
The SNAP reference mission is presented in this paper
in the form of a spacecraft, orbit,
telescope, imager, spectrograph, and observation strategy.
This mission serves as a reference from which future optimization
studies will be performed and
is a demonstration of a design that can
realistically accomplish the science goals.
It is self-consistent,
technically feasible, reduces risk, and is specifically designed to
provide data that strictly control statistical and systematic uncertainties
for supernova and weak gravitational lensing science.

SNAP can run as a complete supernova experiment providing discovery,
and photometric and spectroscopic followup of a full sample of
supernovae over the target redshifts. SNAP is also an excellent
platform for performing deep wide-field surveys with space-quality
imaging
providing crucial complementary constraints to study dark energy
from weak lensing and other studies.
SNAP's full potential is realized when used in conjunction with
complementary results that are anticipated by the time SNAP launches:
a large sample of well-measured low-redshift SNe Ia and precise
measurements of the distance to the surface of last scattering from the Planck satellite.

In \S\ref{dataset:sec} we describe the precision
cosmology that can be achieved using Type Ia supernovae.
The 
needed supernova data set
is based on the
identification of known or plausible
systematic uncertainties that fundamentally limit
measurement accuracy, and on the number and redshift range of supernovae
necessary to probe precisely the dark-energy parameters.  The
power of weak gravitational lensing surveys in measuring
cosmological parameters is discussed in
\S\ref{lensing:sec} and the added complementarity of lensing
and other dark-energy probes made possible with SNAP data
is described in \S\ref{otherde:sec}.  We present in
\S\ref{proposed_experiment} the observing strategies and instrumentation 
suite tailored to provide the data that satisfy both the statistical
and systematic requirements for the supernova and lensing surveys.
The SNAP calibration program is described in \S\ref{calib:sec}.
The results of detailed simulations of the SNAP mission are
given in \S\ref{sim:sec}.
In
\S\ref{ancsci:sec} we discuss the general properties of the SNAP
surveys and the science resources they will provide.

\section{Cosmology with Supernovae}
\label{dataset:sec}

Type Ia SNe have already
proved to be excellent distance indicators for probing the dynamics of
the Universe.
However, the current supernova cosmology measurements are
limited by systematic uncertainties (which occur for all cosmological probes).  In order to move toward the era of
precision cosmology, a new data set must both address
potential systematic uncertainties and provide sufficient statistical
power to accurately determine dark-energy properties,
especially the time variation of the
equation of state.  
Because control of systematic uncertainties is so central
we begin with aspects of the data set necessary for their control.
We proceed by listing the sources of systematic
uncertainty, strategies to reduce their effects, and the resulting
improvements that can be obtained in measuring the supernova Hubble diagram.
We then determine which supernova
observations give a statistical uncertainty comparable to the systematic
uncertainty, leading to
requirements on the number of supernovae we need to find, their distribution
in redshift, and how precisely we need to determine each one's peak
brightness.

\subsection{Control of Systematic Uncertainties}
\label{exec_uncertain}
 
\subsubsection{Known Sources of Systematic Uncertainties}
High-redshift supernova searches have been
proceeding since the late 1980's \citep{norgaard:1989,couchetal:1989,
perlmutteretal:1995,7results97,nature98,
42SNe_98,schmidt_98,riess_acc_98,tonryetal:2003,knopetal:2003,
riessetal:2004}.
Particularly since the discovery of the accelerating expansion of the Universe,
the high-redshift supernova methodology for measuring cosmological
parameters
has been critically scrutinized for sources of systematic uncertainty.
Below are identified systematic effects which any experiment aiming to
make maximal use of the supernova technique will need to recognize and
control.
We also give a rough
estimate of the expected {\it systematic} residual in supernova
magnitude after statistical correction for such effects with the SNAP
data set or from independent data sets from other astrophysical studies.

\noindent
{\it Extinction by Host-galaxy ``Normal'' Dust:} Extinction from
host-galaxy dust can significantly reduce the observed brightness of a
discovered supernova although typically Type Ia supernovae
suffer only modest extinction \citep{hatano98}.  Cross-wavelength
flux-calibrated spectra and multi-band photometry will identify the
properties of the obscuring dust and gas and the amount of extinction
suffered by individual supernovae.  There will remain a residual
uncertainty proportional to the calculated extinction due to dust-model
dependence.  This can be minimized by excluding highly-extincted
supernovae from the analysis.  Stringent requirements on the SNAP
calibration system are needed to control uncertainties in the
photometric-system zeropoints, particularly since the uncertainties
are correlated from supernova to supernova.  Our calibration program
is being designed to control extinction uncertainty to 1\%.

\noindent
{\it Gravitational Lensing by Clumped Mass:} Inhomogeneities along the
supernova line of sight can gravitationally magnify or demagnify the supernova flux
and shift the mode of the supernova magnitude distribution by $\sim$1
-- 10\% depending on redshift
\citep{hw98,mortsell:compact,amanullah:2003}.
  Since flux is conserved in this process, averaging large
numbers of supernovae per redshift bin will give the correct mean brightness.
Deviations from a Gaussian distribution for the brightness of
supernovae at similar $z$ can be used to determine the clumpiness of
the intervening mass distribution
\citep{klmm:03}.  The large SNAP supernova
sample will reduce the distance modulus  uncertainty per redshift bin to
$\sim0.5$\%.  SNAP weak gravitational lensing
measurements, and micro-lensing studies can further help distinguish
whether or not the matter is in compact objects.

\noindent
{\it K-Correction and Cross-Filter Calibration:} Broadband photometry
of supernovae at different redshifts is sensitive to differing
supernova-rest-frame spectral regions.
K-corrections are used to put these
differing photometry measurements onto a consistent rest-frame
passband \citep{kim_kcorr96,Nugent:2002}.
Applying K-corrections with only today's methods and calibration
sets could produce errors of several percent.   Using the currently available spectra,
we tune the SNAP filter-set specifications
to give the target systematic uncertainty $<0.02$ mag.
(The situation will improve even further as the spectral
library of supernovae grows and SNAP itself will add
to this library.)  Furthermore, strict requirements are
placed on the SNAP calibration program to ensure accurate transformations
of fluxes between filters.

\noindent
{\it Galactic Extinction:} Extinction maps of our own Galaxy
are uncertain by $\sim$ 1--10\% depending on direction
\citep{schlegel_dust_98}. Supernova fields can be chosen toward the
low extinction Galactic caps. Future Spitzer observations will allow an
improved mapping between color excesses (e.g.\ of Galactic halo subdwarfs
in the SNAP field) and Galactic extinction by dust.
Galactic-extinction uncertainty can then be controlled
to $<0.5$\% in brightness.

\noindent
{\it Non-SN Ia Contamination:}  Other supernova types are on average
fainter than SNe Ia and their contamination could bias their
Hubble diagram.  Observed supernovae must be positively
identified as SN~Ia.  As some Type~Ib and Ic supernovae have spectra and
brightnesses that otherwise mimic those of SNe~Ia, a spectrum covering
the defining rest frame Si~II 6150\AA\ feature for every supernova at
maximum will provide a pure sample.  This systematic would then be eliminated.

\noindent
{\it Malmquist Bias:} A flux-limited sample preferentially detects
the intrinsically brighter members of any population of sources.
The amount of magnitude bias depends on details of the search but
can reach the level of the intrinsic magnitude dispersion of a standard
candle.
Directly correcting this bias would rely on knowledge of the SN Ia
luminosity function, which may change with lookback time.  A detection
threshold fainter than peak by at least five times the intrinsic SN~Ia
luminosity dispersion ensures sample completeness with respect to
intrinsic supernova brightness, eliminating this bias \citep{klmm:03}.

\subsubsection{Possible Sources of Systematic Uncertainties}
A systematics-limited experiment must
account for speculative but reasonable sources of uncertainty.  
The following are sources of systematic uncertainty for which there is no
direct evidence but which cannot yet be discounted.

\subsubsubsection*{Extinction by Gray Dust} As opposed to normal dust,
gray dust is postulated to produce wavelength-independent absorption
in optical bands \citep{aguirre_99,aguirrehaiman:2000}.  Although
models for gray-dust grains dim blue and red optical light equally,
the near-infrared (NIR) light ($\sim$1-2 $\mu$m) is less affected.
Cross-wavelength calibrated spectra extending to wavelength regions
where ``gray'' dust is no longer gray will characterize the
hypothetical large-grain dust's absorption properties.  Armed with the
extinction -- color excess properties of the gray dust, broadband
near-infrared colors can provide ``gray'' dust extinction corrections
for supernovae out to $z=0.5$; \citet{Goobar:2002} have explored the
observations necessary to be sensitive to 0.02 mag gray-dust
extinction.

Gray dust will re-emit absorbed starlight and thus contribute to
the far-infrared background; current observations indicate that the FIR
flux is attributable to point sources
\citep{Borys:2002,scott:2002}.  Quasar colors in
the SDSS sample as a function of redshift give an upper limit to the
possible dimming due to gray dust at the level of 0.2 mag (in
the restframe B-band) at $z=1$ \citep{mortsell:2003}.
Deeper SCUBA and Spitzer observations should further tighten the
constraints on the amount of gray dust allowed.



\subsubsubsection*{Uncorrected Supernova Evolution}
Supernova behavior itself may have systematic variations depending,
for example, on properties of the progenitor systems.  The
distribution of these stellar properties is likely to change over
time---``evolve''---in a given galaxy, and over a population of
galaxies.
Supernova heterogeneity manifests itself through observed signatures
in their spectral light emission (the bulk emitted in optical
wavelengths), and the time-evolution of that emission.  Signs of 
supernova diversity are thus captured in observations of  multi-band
light curves and/or spectral time-sequences
that begin shortly after explosion and extend to the nebular phase
of the event.  This has been done
using optical measurements of
nearby SNe~Ia drawn from a wide range of galactic
environments that provide an observed evolutionary range of SNe~Ia
\citep{hamuy96,hamuy00}.  The photometric and
spectral differences that have been identified
in these data are well calibrated by the SN~Ia light curve
width-luminosity relation, leaving a 10\%
intrinsic peak-brightness dispersion.
There is currently no evidence for systematic residuals after
correction, for example with galaxy type or supernova location
\citep{sullivan:2002}.

High signal-to-noise ($S/N$) multi-band light curves and spectra over
optical wavelengths can provide precise control over additional
general supernova variability.  Efforts are underway to
expose additional effects using larger, more precise and
systematic, low-redshift supernova surveys \citep{aldering02}.
These high-quality data are also necessary to control possible evolution
effects when performing precision cosmology with high-redshift supernovae.

Theoretical models can identify observables that are expected to
display heterogeneity.  These key features, indicative of the
underlying initial conditions and physical mechanisms controlling the
supernova, will be measured with SNAP, allowing statistical correction for
what would otherwise be a systematic uncertainty.  The state of empirical
understanding of these observables at the time SNAP launches will be
explicitly tested by SNAP measurements \citep{branchetal:2001}.
At present, we perform Fisher
matrix analyses on model spectra and light curves to estimate the
statistical measurement requirements, and to ensure that we have
necessary sensitivity to use subsamples to test for residual
systematics at better than the 2\% level. This approach reveals the
main effects of --- as well as the covariance between --- the following
observables:



\noindent
{\it Rise time from explosion to peak:} This is an
indicator of opacity, fused $^{56}$Ni mass, and possible differences
in the $^{56}$Ni distribution. A 0.1~day uncertainty corresponds to
a 1\% brightness constraint at peak \citep{hoflich98}.  Achieving
such accuracy requires discovery within $\sim$2 days of explosion,
on average, $\sim$30$\times$ fainter than peak brightness.  Current
constraints on rise-time differences are two days
\citep{alderingrise}.

\noindent
{\it Plateau level 45 days past peak:} The light-curve plateau level
that begins $\sim$45 days past --- and more than 10$\times$ fainter
than --- peak is an important indicator of the C/O ratio of the
progenitor star, and fused $^{56}$Ni.  In models, a 5\% constraint on this
plateau brightness corresponds to a 1\% constraint on the peak
brightness \citep{hoflich98}.


\noindent
{\it Overall light-curve timescale:} The ``stretch factor''
(\citet{goldhaber:2001}, see also \citet{rpk96,7results97,phillips99})
that
parameterizes the light-curve timescale \citep{phillips:1993} is affected by almost all of
the aforementioned parameters since it tracks the SN~Ia's light-curve
development from early to late times.  It is correlated with rise time
and plateau level and it ties SNAP's controls for systematics to the
controls used in the current ground-based work.  A 0.5\% uncertainty in
the stretch factor measurement corresponds to a $\sim$1\% uncertainty
at peak \citep{42SNe_98}.  Stretch distributions are quite consistent
between the current sets of low and high-redshift supernovae
\citep{42SNe_98,knopetal:2003}.

\noindent
{\it Spectral line velocities:} The velocities of several spectral
features throughout the UV and visible make an excellent diagnostic of
the kinetic-energy injection in SNe~Ia. In models, velocities constrained to
$\sim$250~km~s$^{-1}$ constrain the peak luminosity to $\sim$1\%
\citep{hoflich98}, given a typical SNe~Ia expansion velocity of
15,000~km~s$^{-1}$.  Current data show smooth velocity development for
a given supernova, but also clear differences between supernovae which have
not yet been found to correlate with supernova luminosity
\citep{bvdb93,hatanoetal:2000}.

\noindent
{\it Spectral features:} The positions of various spectral features in
the restframe UV are strong metallicity indicators of the SNe~Ia.  By
achieving a sufficient $S/N$ and resolution on such features \citep{bk:03} SNAP
will be able to constrain the metallicity of the progenitor to 0.1 dex
\citep{lentz00}.  This spectral region has only recently begun to be
explored with UV spectroscopy of nearby supernovae with HST.  Spectral
features in the restframe optical (Ca~II H\&K and Si~II at 6150 \AA)
provide additional constraints on the opacity and luminosity of the
SN~Ia \citep{nugent95}.

\begin{figure}[h]
\plotone{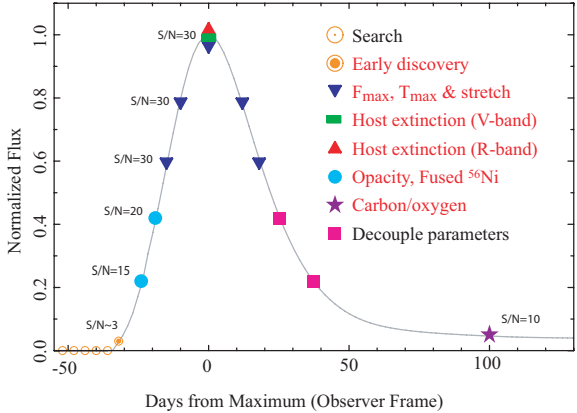}
\caption{Sample B-band light curve for a $z=0.8$ Type Ia supernova.
Signal-to-noise targets are shown at different epochs for identification
of possible systematic effects.
\label{syslc:fig}}
\end{figure}

\begin{figure}[h]
\plotone{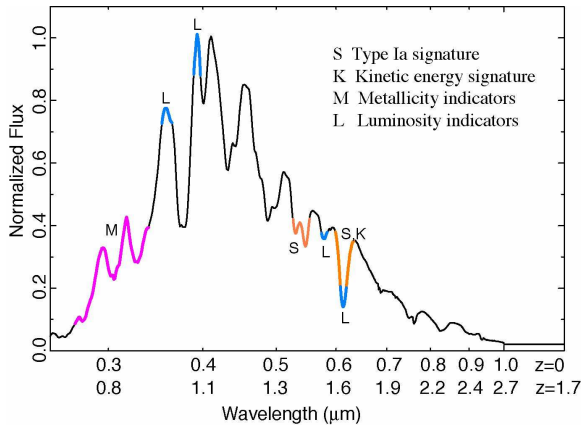}
\caption{Type Ia spectra at maximum light showing line features
associated with possible Type Ia supernova variation.  The
horizontal axis shows observer frame wavelengths for $z=0$ and 
$z=1.7$ supernovae.
\label{sysspec:fig}}
\end{figure}

Figures~\ref{syslc:fig} and \ref{sysspec:fig} show the particular
light-curve and spectral parameters that can serve as indicators of
supernova evolution.  By measuring all of the above features for each
supernova we can tightly constrain the physical conditions of the explosion. 
This makes it possible to recognize subsets of supernovae with matching initial
conditions, ensuring a small luminosity range for each subset.

%

In addition to these features of the supernovae themselves, we will
also study the host galaxy of the supernova.  We can measure the host
galaxy luminosity, colors, morphology, and the location of the supernova
within the galaxy, even at redshifts $z\sim1.7$.  The last
two observations
are difficult or impossible from the ground for high-redshift
galaxies.  These galaxy properties can provide clues to any
shifts with redshift in the supernova progenitor population.

\subsubsection{Systematic-Error Model}
\label{syserrmod:sec}
The behavior of systematic uncertainties are complicated and the
determination of their propagated effects on cosmology
requires rigorous analysis
\citep{klmm:03,hutererkim:2004}.  
Particularly important systematic effects for supernova-cosmology
experiments are those that put a floor on how well the distance
modulus can be measured in a certain redshift bin.  For example,
supernovae at similar redshifts whose rest-frame flux regions lie in
common observer passbands can share completely correlated calibration
uncertainty.  An increase in the statistical sample of supernovae
cannot decrease this uncertainty.  In addition, these
uncertainties could be correlated across redshift bins.  Again taking
calibration as an example, a temperature uncertainty in a blackbody
fundamental calibrator propagates into correlated flux and magnitude
zeropoint uncertainties.  Correlated errors across redshift bins can
lead to a bias in the determination of the cosmological and
dark-energy parameters.  We do not further consider correlated bias
in this paper as \citet{klmm:03} find that the systematic bias in
$w_0$ and $w'$ lies well within the 68\%-error contours due to
comparably-sized irreducible systematic uncertainties.

For convenience we adopt
in this paper a systematic-uncertainty model of
$0.02(1.7/z_{max})(1+z)/2.7$ 
magnitude irreducible uncertainty per
redshift bin of width 0.1 as proposed by \citet{hut:2003}, 
where $z_{max}$ represents the depth of the
survey. The systematic uncertainty introduced in the comparison of
high-redshift supernovae with the local sample 
is an increasing function of redshift.
Systematic uncertainty will
exist at some level even for the local sample, due for example to
uncertainty in the Galactic-extinction zeropoint and spatial
correlations in $E(B-V)$ maps.

\subsection{Comparing Defined Subsets of Supernovae}
The subtyping strategy employed for evolution based on 
observables tied to the explosion characteristics can be 
more generally applied. 
The data (supernova risetime, early detection, light-curve peak-to-tail ratio, identification
of the Type Ia-defining Si II spectral feature, separation of supernova
light from host-galaxy light, and identification of host-galaxy
morphology, etc.) make it possible to study each individual supernova
and measure enough of its physical properties to recognize deviations
from standard brightness subtypes.  Only the change in brightness as a
function of the parameters classifying a subtype is needed, not any
intrinsic brightness. Supernovae cannot change their brightness in one
measured wavelength range without affecting brightness somewhere else in
the spectral time series ---  an effect that is well-captured by
expanding atmosphere computer models.

As a precision experiment SNAP is designed to distinguish supernovae
``demographics'' by detecting  subtle variations in
light-curve and spectral parameters. 
By matching
like to like among the supernova subtypes we can construct
independent Hubble diagrams, each with systematic biases
reduced to the level of 1\% in distance.
This procedure is described and tested in \citet{klmm:03}. 
Comparison within a redshift bin reveals possible systematics, 
while comparison within a subtype over the redshift range cleanly 
probes the cosmology. 

The targeted
residual systematic uncertainty from effects such as Malmquist bias,
$K$-correction, etc.~total
$\sim$2\%. Thus, a subset, or ``like vs.~like'', analysis based on 
physical conditions should group supernovae to within $<$2\% in luminosity. 
This is in addition to a purely statistical correction for any 
second-parameter effects beyond the stretch factor.

\subsection{Supernova Sample to Probe Dark Energy}
\label{asprobe}
The statistical supernova sample appropriate for experiments with the
presence of systematic uncertainty has been considered in the
Fisher-matrix analysis of \citet{hut:2003}. (Their results have been
validated with our independent Monte Carlo analysis.) They assign
a statistical
uncertainty of 
0.15 magnitudes to each supernova, which combines
statistical measurement uncertainty and intrinsic supernova
dispersion.  The systematic-error model described in
\S\ref{syserrmod:sec} is included.

\begin{figure}[h]
\plotone{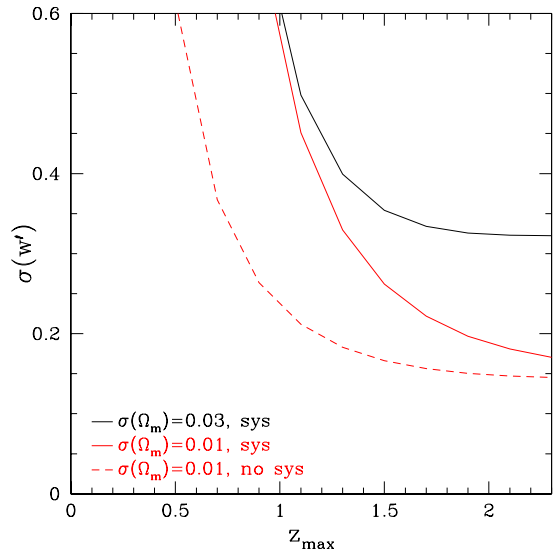}
\caption[Parameter estimation as a function of maximum redshift.]
{From \citet{hut:2003}, accuracy in estimating the equation of state variation parameter, $w'$, as a
function of maximum redshift probed in supernova distance surveys.
The lower two curves assume priors in $\Omega_M$ good to 0.01, 
while the upper two
curves are for the case where systematic uncertainties are present. The top, heavy curve corresponds to the most realistic case.
It is clear that even with modest systematic uncertainties good accuracy
requires probing to high redshift.  In all cases a flat universe is
assumed.  \label{sci_deltaw.fig}}
\end{figure}

The importance of using SNe~Ia over the full redshift range out to
$z\sim 1.7$ for measuring the cosmological parameters is demonstrated
in Figure~\ref{sci_deltaw.fig}, which shows the uncertainty in
measuring the equation of state parameter variation, $w'$, 
as a function of maximum redshift probed in distance
surveys \citep{hut:2003}.  (Note that the results look very similar 
whether one defines $w'$ in terms of a linear expansion of $w(z)$ in 
redshift or in scale factor, and whether one uses a prior on matter 
density or on distance to the CMB last scattering surface.)  
This calculation considers 2000 SNe~Ia
measured in the range $0.1\le z\le z_{\rm max}$, along with 300
low-redshift SNe~Ia from the Nearby Supernova Factory
\citep{aldering02}.  A flat universe is assumed.  Three types of
experiments were considered. The first is an idealized experiment
subject only to statistical uncertainties, free of any systematics,
and with extremely tight prior knowledge of the matter density. The
second two are more realistic models which assume both statistical and
systematic uncertainties and different priors on $\Omega_M$.

From this figure we conclude that a SN~Ia sample extending 
to redshifts
of $z>1.5$ is crucial for realistic experiments in which some 
systematic uncertainties remain after all statistical corrections are
applied. 
Ignoring systematic uncertainties can lead to
claims that are too optimistic.  Similar conclusions apply for the other
dark-energy parameters, including $w_0$ and an assumed constant $w$
\citep{hut:2003}.

The sample of $\sim$2000 SNe~Ia is enough to provide
systematic-uncertainty limited measures of the dark-energy parameter
$w'$.  These supernovae represent a subset of SNe Ia that can provide
robust cosmological measurements (e.g.\ non-peculiar, low-extinction),
culled from an even larger set of discovered supernovae.  This large
sample is necessary to allow model-independent checks for any residual
systematics or refined standardization parameters, since the sample
will have to be subdivided in a multidimensional parameter space of
redshift, light curve-width, host properties, etc.  The large number
of lower signal-to-noise light curves from supernovae at $z>1.7$ will
provide an extended redshift bin in which to check for consistency
with the core sample.

Expected parameter measurement precisions for the SNAP supernova
program with $z_{max}=1.7$ are summarized in
Table~\ref{sci_errors.tab}.  We determine precisions for two fiducial
flat universes with $\Omega_M=0.3$, one in which the dark energy is
attributed to a Cosmological Constant and the other to a
SUGRA-inspired dark-energy model.  The parameter precisions then
depend on the choice of data set, priors from other experiments,
assumptions on the flatness of the universe,
and the model for the behavior of $w$.
(Parameter determination from Monte Carlo
analysis of simulated data from the reference SNAP mission is given
in \S\ref{sim:sec}.)

\begin{deluxetable}{cccccc}
\tablewidth{0pt}
\tablecolumns{5}
\tablecaption{SNAP 1-$\sigma$ uncertainties in dark-energy parameters, with conservative systematics for the supernova and a 1000 sq.\ deg.\ weak-lensing survey.  Note: These uncertainties are systematics-limited, not statistics limited.\label{sci_errors.tab}}
\tablehead{
& $\sigma_{\Omega_M}$ & $\sigma_{\Omega_w}$ & $\sigma_{w_0}$ &
$\sigma_{w'}$
}
\startdata
Fiducial Universe: flat, $\Omega_M=0.3$, Cosmological Constant dark energy \\
\cline{1-1}\\
SNAP SN; $w=-1$& 0.02 &  0.05 & \nodata & \nodata  \\
SNAP SN + WL; $w=-1$ 
                       & 0.01  &  0.01   & \nodata &\nodata\\
SNAP SN;$\sigma_{\Omega_M}=0.03$ prior; flat; $w(z)=w_0+2w'(1-a)$ 
                       & \nodata &  $0.03$ & $0.09$ & $0.31$  \\
SNAP SN;Planck prior; flat; $w(z)=w_0+2w'(1-a)$ 
                       & \nodata  &  $0.01$   &$0.09$ & $0.19$  \\
SNAP SN + WL; flat; $w(z)=w_0+2w'(1-a)$ 
                       & \nodata  &  $0.005$   & $0.05$&$0.11$\\
\\
Fiducial Universe: flat, $\Omega_M=0.3$, SUGRA-inspired dark energy\\
\cline{1-1}\\
SNAP SN; $\sigma_{\Omega_M}=0.03$ prior; flat; $w(z)=w_0+2w'(1-a)$ 
                       & \nodata  &  $0.03$   & $0.08$&$0.17$\\
SNAP SN; Planck prior; flat; $w(z)=w_0+2w'(1-a)$ 
                       & \nodata  &  $0.02$   & $0.09$&$0.13$\\
SNAP SN + WL; flat; $w(z)=w_0+2w'(1-a)$ 
                       & \nodata  &  $0.005$  & $0.03$&$0.06$\\
 
\enddata
\tablecomments{Cosmological and dark-energy parameter
precisions for two fiducial
flat universes with $\Omega_M=0.3$, one in which the dark energy is
attributed to a Cosmological Constant and the other to a
SUGRA-inspired dark-energy model.  The parameter precisions then
depend on the choice of data set, priors from other experiments,
assumptions on the flatness of the universe,
and the model for the behavior of $w$.
In this paper we adopt the non-standard
definition $w'\equiv dw/d\ln a|_{z=1}$}
\end{deluxetable}

If the universe is taken to be spatially flat, the SN data (with
Planck prior) determine $\Omega_M$ to 0.01.  The present value of the
dark energy equation of state is constrained within 0.09 and the
physically crucial dynamical clue of the equation of state time
variation is bounded to within 0.19.
Note that these constraints are {\it systematics limited};
much tighter constraints would be obtained if systematics were
ignored.  (In \S\ref{otherde:sec}  we discuss the tighter constraints obtained by
combining the supernova results with the other SNAP measurements.)

The
precise, homogeneous, deep supernova data set allows for robust
interpretation in more general terms than $w_0$ and $w'$. The function
$w(z)$ can be implemented in a nonparametric, uncorrelated bin, or
eigenmode method \citep{HutererStarkman:2003,HutererCooray:2004}.  One
can also obtain the expansion history $a(t)$ itself, e.g.\
\citet{linder:2003}.

We can also obtain constraints on flatness of the universe, rather
than assuming it. Current data are mostly sensitive to one linear
combination of $\Omega_M$ and $\Omega_{\Lambda}$ (roughly their
difference). SNAP is designed to obtain sufficient brightness-redshift
data for a wide range of redshifts ($0.1<z<1.7$), and to enable an
extraordinarily accurate measurement of $\Omega_M$ and
$\Omega_{\Lambda}$ separately (see Figure~\ref{confcmbclust}), the
culmination of an approach originally proposed by \citet{goobar95}.
Assuming that the dark energy is the cosmological constant, the
supernova experiment can simultaneously determine the mass density
$\Omega_M$ to an accuracy of $0.02$, the cosmological constant energy
density $\Omega_{\Lambda}$ to $0.05$.  Propagating this through one
obtains a 0.06
measurement uncertainty of the curvature of the universe,
$\Omega_k=1-\Omega_M-\Omega_{\Lambda}$.

\section{Cosmology with Weak Lensing}
\label{lensing:sec}
Gravitational weak lensing provides an independent and complementary
measurement of the dark-energy parameters through the mapping of
galaxy-shape distortions induced by mass inhomogeneities in the
Universe.  Weak gravitational lensing of background galaxies by
foreground dark matter (large-scale structure) provides a direct
measurement of the amount and distribution of dark matter (see
\S\ref{science:sec} for details on the impact SNAP can have on these
measurements). The growth of these dark matter structures with cosmic
time is determined by the dark energy.  Thus, precise weak lensing
observations can provide significant and independent constraints on
the properties of dark energy.  For a review of the techniques used
to measure weak lensing and the current status of weak-lensing
measurements, see
\citet{refregier:2003}.

\subsection{Weak Lensing as a Probe of Dark Energy}

Dark energy modifies the weak-lensing observables by altering the
distance-redshift relation as well as the matter power spectrum. These
two, in turn, determine the weak-lensing shear power
spectrum \citep{weak_kaiser_98}.  
Specifically, dark energy will modify the normalization and nonlinear
part of the matter power spectrum and
redshift evolution.  Weak
lensing measurements with a  wide-field space telescope such as
SNAP can thus achieve an accurate measurement of $\Omega_w$,
and the dark-energy parameters
\citep{hu:1999,huterer:2002,abazaj:2003,refregieretal:2003}.

\begin{figure}
\plotone{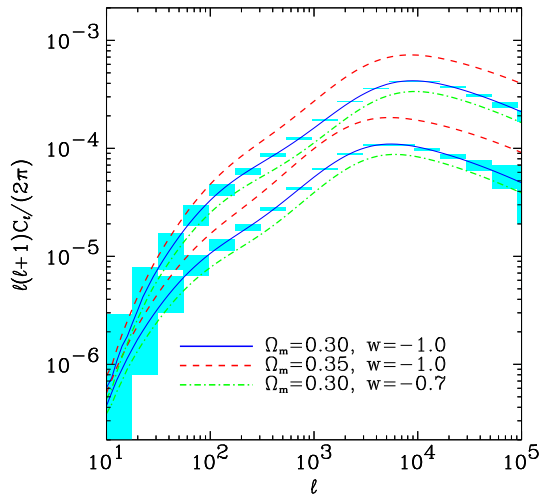}
\caption{The dark matter power
spectrum in two redshift intervals. The upper curve represents a
measurement made with lens galaxies at $z>1$ and the lower curve a
measurement with $z<1$.  Three different cosmological models are
shown.  It is clear that the over 100 million resolved galaxies in
the SNAP wide survey will distinguish between cosmological models
even after being divided into redshift intervals.  Therefore, high
signal-to-noise measurements of the evolution of the power
spectrum will be possible.
\label{power:fig}}
\end{figure}

\begin{figure}
\plotone{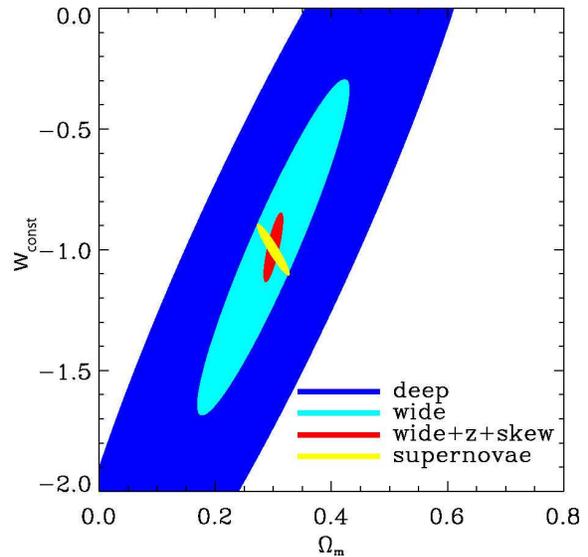}
\caption{Joint constraints on $\Omega_M$ and $w_{const}$ achievable with the
SNAP deep and wide surveys using weak gravitational lensing.  The blue
contour is for the SNAP supernova survey and cyan contour for the
wide-field survey.  After dividing the wide survey into two redshift
intervals and adding skewness information, the constraints (the red
contour) are similar in size to the SN Ia constraints (shown as a
yellow ellipse).  Since weak-lensing constraints are largely
orthogonal to those of SNe Ia, weak lensing plays a crucial role in
SNAP's ability to measure dark energy.
\label{snapconst:fig}}
\end{figure}

In Figure~\ref{power:fig} we show the expected measurement of the
shear power spectrum possible with 100 million galaxies over 300
square degrees divided into 2 redshift bins. The upper curve
represents a measurement using resolved source galaxies with $z>1$
while the lower curve represents $z<1$ galaxies. The power of
space-based observations is most apparent at small scales (high
$\ell$), where the shot noise is small due to the large surface
density of resolved galaxies. In Figure~\ref{snapconst:fig} we show
the joint constraints on $\Omega_M$ and assumed-constant $w$ using 10
million galaxies, 100 million galaxies,
and 100 million galaxies divided into two redshift bins with
additional skewness information. Note that the constraints from weak
lensing are largely orthogonal to the constraints from
supernovae. Thus, weak lensing complements the supernova technique in
deriving constraints on $w$.

\citet{takada:2003:mnras} have demonstrated that by
combining additional information on non-Gaussianity from the lensing
bispectrum in the moderately nonlinear regime with the power spectrum,
cosmological constraints can be improved significantly (about a factor
of three over the results shown in Figure~\ref{snapconst:fig}).  At
present the ability to exploit this information is limited by
computationally intensive numerical predictions for non-linear growth,
but the gravitational physics involved is well understood and work on
obtaining accurate predictions is already underway.

\begin{figure}
\plotone{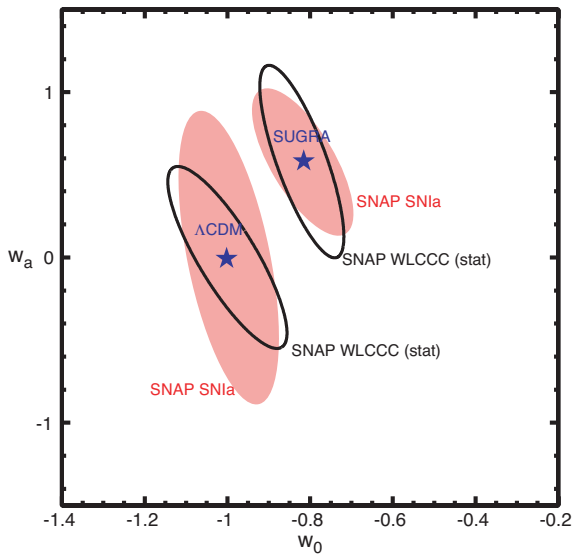}
\caption{Constraints on the parameters $w_0$ and $w_a \approx 2w'$, where
$w(z)=w_0+w_a(1-a)$ and $a$ is the scale factor \citep{linder:2003}.
Shown are the SNAP constraints from the supernova survey (filled
contours), and those from weak-lensing cross-correlation cosmography
(empty contours; statistics only), as discussed in the text.  We show constraints for
two models: vacuum energy (``$\Lambda$CDM'') and a supergravity-inspired model
(``SUGRA'') \citep{Brax:1999}. 
\label{bj03fig}}
\end{figure}

At still smaller (arcminute) angular scales, the lensing signal is
very strong, but the non-linear growth predictions become very difficult.
New statistical treatments of weak-lensing data have recently been
proposed
\citep{jaintaylor:2003,bernsteinjain:2003} which measure the cosmic geometry
while canceling uncertainties in the non-linear growth.  This is
achieved by cross-correlating the lensing signal with an estimate of
the foreground mass distribution.  Figure~\ref{bj03fig} presents an
estimate by \citet{bernsteinjain:2003} of the statistical uncertainties
on $w_0$ and $w_a$ (or $w'$, recall $w_a \sim 2w'$ at $z\sim 1$)
that could be obtained by the SNAP surveys using this
lensing cross-correlation technique.  The quality of the measurement
is comparable to that of the Type Ia SN measurements, as long as
the systematic uncertainties are kept below the statistical uncertainties.

\subsection{Advantages of Space-Based Measurements}
Several conditions are necessary to make possible high signal-to-noise
weak-lensing measurements for these types of analyses.  The shapes of
millions of background galaxies must be measured to overcome cosmic
variance and provide a precise measure of the dark matter power
spectrum.  It is particularly important to be able to resolve many
small, distant galaxies beyond a redshift of $z\simeq 1$; the broad
redshift range of source galaxies provides a powerful lever arm on the
constraint of $w$ by providing information on the growth of
structure. High spatial resolution is necessary to extract
statistically powerful shape information from each individual galaxy
and decrease the size of the smallest resolved galaxies in a survey.
A stable point spread function (PSF) and a minimal level of internal
optical distortions are essential for the precise measurement of
galaxy shapes. The high spatial density of resolved background
galaxies afforded by space-based observations allows
cross-correlation measurements and gives measurements of the mass
power spectrum at scales where non-linearities are most apparent.
Accurate photometric redshifts are required for the lensing analyses,
particularly the lensing cross-correlation, where 
any systematic biases in photo-z's above $\sim$0.001 in
$\log{(1+z)}$ would
dominate the error budget.  Catastrophic photo-z errors would similarly
need to be kept below 1 per 1000.  It may be possible use internal
analyses of lensing data to trade precision on the dark energy
parameters for looser constraints on photo-z biases.  Reducing the
photo-z errors, biases, and catastrophic errors requires high-accuracy
photometry in bands spanning the visible and near-IR.  Atmospheric
emission lines make it impractical to acquire these data in the near-IR
and near certain bright visible lines.

Satisfying all of these requirements simultaneously requires a
wide-field space based imaging telescope in a thermally stable orbit.
With a high-throughput, space-based 2 meter telescope, an exposure
time of 2000s will provide the depth to resolve significant numbers of
galaxies in the scientifically interesting range beyond $z\simeq
1$. Unhindered by atmospheric emission and absorption, NIR
observations from space provide significant redshift depth and
improved photometric redshift estimates.  Avoiding the Earth's
atmosphere in a thermally stable orbit gives the fine resolution and
stable PSF.  The PSF and depth give a high galaxy surface density.  A
ground-based survey of tens of thousands of square degrees, exploring
the mass power spectrum on large angular scales with small statistical
errors would be complementary. In particular, such large-angular-scale
surveys probe the linear part of the matter power spectrum while
deeper but somewhat less wide surveys probe the nonlinear region.
(For completeness, we discuss the possibility of a panoramic,
wide-field survey with SNAP in \S\ref{panoramic:sec}.)

\section{Joint SNAP Constraints on Dark Energy from Weak-lensing, Supernova,
and other Measurements}
\label{otherde:sec}
The two complementary cosmological surveys that
constitute the leading components of the SNAP primary mission 
together combined give constraints on the properties
of dark energy significantly stronger than either one by itself.
Table~\ref{sci_errors.tab} gives the
expected dark-energy parameter uncertainties from the joint results
from the deep supernova survey and the combination of power spectrum,
bispectrum, and cross-correlation cosmography techniques from a 1000
square-degree lensing program with $n_{gal}=100$ arcmin$^{-2}$.  A
very conservative estimate of the systematics floor on shear
calibration of 3\% is used.

For a flat universe the present dark
energy equation of state $w_0$ can be measured to $\pm 0.05$, and
the time variation can be bounded within 0.11.  Without assuming a
flat universe,
the mass density
$\Omega_M$, the cosmological constant energy
density $\Omega_{\Lambda}$, and curvature $\Omega_k$
can be determined to $\ls 0.01$. 
This poses an interesting test of the cosmological
framework by cross-checking the curvature measured at $z \sim 1$ with
current measurements out to $z \sim 1000$ of $\Omega_k=-0.02\pm 0.02$
\citep{bennettetal:2003}.
Note
that {\it no priors} are required when we consider the full SNAP
mission of supernovae plus weak lensing.  These are significant
improvements over the results from supernovae alone.  The constraints
are subject to improvement as practical understanding of future weak
lensing data develops.

Universes with more dark energy (than the fiducial $\Omega_{DE}=0.7$
assumed here), dark energy possessing a more positive equation of
state or a stronger time variation would lead to a measurement with
more sensitivity to dark-energy properties.  We have been conservative
in taking the cosmological constant as fiducial, the least sensitive
of the canonical quintessence models.  For example, in a
supergravity-inspired model of dark energy \citep{Brax:1999}, $w'$
would be more tightly constrained to $\pm 0.06$.
With strong constraints on $w'$ we will be able to differentiate
between the cosmological constant and a range of dynamical
scalar-field (``quintessence'') particle-physics models
(\citet{weller01,weller02,hutererturner:2001}.)

The SNAP reference mission focuses on the established methods of Type
Ia supernova Hubble diagrams and weak gravitational lensing to study
dark energy.  Other less-tested and more model-dependent methods for
measuring cosmological parameters are now being explored as additional
probes of dark energy with SNAP.  These include (but are not limited
to) using Type II supernovae as distance indicators
\citep{kirshner:1974,b93j1,b93j3,leonard:2003}, baryon oscillations as a
standard ruler in both transverse and radial directions
\citep{Seo:2003}, redshift-space correlations for the Alcock-Paczynski test
\citep{Matsubara:2003}, strong lensing time-delays and
statistics \citep{holz:2001, goobaretal:2002}, and cluster-counting
statistics \citep{haiman:2001}.  Attacking dark energy with
measurements from independent techniques that are subject to
differing, but tightly controlled systematic uncertainties will give
increased confidence in the ultimate conclusions made about its
nature.  Since these techniques generally constrain different combinations
of cosmological parameters, it is important that each ``stand on
its own two feet'' with systematics control so that they can be
combined for studies of dark energy.

SNAP can naturally study dark energy using many of these other probes.
The necessary measurements can be made with the SNAP instrumentation
suite as is, or in coordination with ground or other space-based
telescopes.  A SNAP panoramic-survey observing mode that covers a
significant fraction of sky can provide large number statistics,
redshift depth, wavelength coverage, discovery of rare objects, and
measures on large angular scales; these are exactly the
characteristics of the data needed for many of these complementary
methods, with all the advantages of observations above the distorting
and absorbing atmosphere.

\section{Reference Model SNAP Experiment}
\label{proposed_experiment}

Discovery and study of a larger number
and of more distant supernovae (or any probe) is by itself insufficient
to successfully
accomplish a rigorous investigation of the cosmological
model.  As
shown in \S\ref{exec_uncertain} we must address each of the
systematic concerns while making precise supernova measurements, thus
requiring a
major leap forward in measurement techniques. The requirements
placed on the SNAP instrument derive directly from the science goals.

The primary requirement is to obtain the corrected peak
brightness vs.\ redshift of at least 2000 Type Ia SNe out to a redshift
of $z=1.7$. Identification of Type Ia SNe requires the measurement of
characteristic spectral features near peak
luminosity. Host-galaxy redshift is also determined
spectroscopically.
The corrected peak magnitudes are derived from supernova rest-frame
optical light curves and spectra.  We require a statistical
measurement uncertainty for the
peak magnitude corrected for extinction and shape inhomogeneity
roughly equal to the current intrinsic magnitude
dispersion of supernovae, $\sim 0.1$ mag, and a systematic uncertainty
almost an order of magnitude smaller.

To meet the scientific requirements, the SNAP reference
mission has a large, 0.7 square degree instrumented field of view and
an observing cadence of 4 days, commensurate with the timescale over
which early-epoch supernova light curves change. The imager's
large field-of-view gives a significant multiplex advantage; each
exposure contains $\sim 80$ Type Ia supernovae within 5.4 observer-frame
months from the date of explosion (corresponding to 2 rest-frame
months for $z=1.7$ supernovae).  Observations in multiple
filters yield multi-color rest-frame optical light curves.  A
spectrometer optimized for supernova spectra is allocated observation
time for supernova type-identification, feature measurements, and
spectral-template building.  The telescope aperture is constrained by
the photometric and spectroscopic $S/N$ requirements and the focal
plane must accommodate the large field of view.

The derived requirements for the SNAP reference mission
described here serve as a starting point;
they are subject to change as the mission is further refined.
As other subsidiary approaches to measure dark energy are developed
as discussed in \S\ref{otherde:sec} we can also consider
minor detector additions or
replacements on the focal plane with either extended broad-band
filters or with dispersive elements.  For example, a band in between
the reference filter bands could sharpen the photometric redshift
accuracy for a part of the survey field.  As a general rule, broadband
photometry is significantly faster than slitless spectroscopy by the
ratio $R_{spect}/R_{filter}$ for equivalent $S/N$,  however,
for specific
applications a
detector with a small dispersive element in place of a filter
may be useful.   It could provide spectral
time series of core-collapse supernovae that fall in
the limited field enabling the determination of
their distance.  It could also provide precise redshifts of field sources that
can calibrate photometric redshifts -- especially useful for
tomographic weak lensing, or for the baryon-oscillation method.

\subsection{Reference Model Observation Strategy and Data Package}
The primary SNAP science program consists of a deep survey and a
wide-field survey with individually-designed observing schedules.  The
repeated scans of the deep-survey field provide supernova discovery
and light curves, and when coadded give a high surface density of
background galaxies for lensing surveys.  The wide-field survey is
designed for dark-energy measurements using weak lensing and other
$w$-sensitive techniques. In addition, we explore a possible panoramic
survey that extends the weak-lensing data and provides the general
astronomical community a ``legacy'' survey of space-quality
optical-to-NIR images of a significant fraction of the sky.  The
programs are described here and summarized in Table~\ref{obs:tab}. The
surveys are described by the solid-angle of sky monitored, the
exposure time for each scan of the field, and the number of times the
survey region is scanned.

\begin{deluxetable}{cccccc}
\tablewidth{0pt}
\tablecaption{Reference SNAP Surveys.\label{obs:tab}}
\tablehead{
Program & Solid Angle & Exposure & Cadence & \# Scans\\
&Per Filter (sq.\ deg.)& Per Scan (s)& (days)& }
\startdata 
Deep North& $7.5$ & Optical: $4\times 300$ & 4 & 120\\
 & &NIR: $8 \times 300$ & &\\ 
Wide-field & $300-1000$ & Optical: $6\times 200$ & \ldots & 1 \\
 & &NIR: $12 \times 200$ & &\\
Deep South& $7.5$ & Optical: $4\times 300$ & 4 & 120\\
 & &NIR: $8 \times 300$ & &\\ 
Panoramic& $7000-10000$ & Optical: $6\times 67$ & \ldots & 1\\
 & &NIR: $12 \times 67$ & &\\ 
\enddata
\end{deluxetable}

\subsubsection{Supernova Program}

A simple, predetermined
observing strategy repeatedly
steps through a 7.5 square
degree northern zone with a 4 day cadence continuing
for 16 months.  Later in the mission this strategy is repeated
for a southern survey region.
The fields are selected to be perpendicular to the
Sun, Earth, and Moon and where
natural zodiacal light is near minimum, toward
the north and south ecliptic caps.
This deep-survey strategy allows discovery and automatic followup for
SNe~Ia that explode in those regions.
Every field will be visited every four days for 16
months, with sufficiently
long exposures that all relevant SNe~Ia in the SNAP survey regions will
have significant signal within a few restframe days of explosion.
(supernovae at much higher redshifts on average will be found slightly later
in their light curve rise times although their prior history
is in the data.) The periodic observation of fixed
fields ensures that every supernova at $z < 1.7$ will have its light
curve followed for at least several months in the rest frame as it
brightens and fades.

Sixty percent of the SNAP deep supernova survey
will be spent for the pure photometric
scan; about 120 exposures are possible with a four day cadence to
cover a scan
area of 7.5 square degrees.
Running this survey again in the south yields
another 7.5 square degrees.

The zodiacal light will be the dominant source of background
given SNAP's fields, orbit, and shielding of sun and earthshine.
The cosmic-ray flux\footnote{From http://crsp3.nrl.navy.mil/creme96/}
contamination of 4.5 particles/cm$^2$/s will
make multiple exposures necessary to eliminate significant contamination of
the images.

Forty percent of the SNAP deep survey will be spent doing targeted
spectroscopy of supernovae with parallel imaging
(that is, imaging occurs simultaneously during
these spectroscopic observations).  The resulting images will cover
random positions and orientations within the SNAP field and will be
used to increase the depth of the survey and help with photometric and
astrometric cross-calibration between detectors.

This prearranged observing program will provide a uniform,
standardized, calibrated data set for each supernova, allowing for the
first time comprehensive comparisons across complete sets of SNe~Ia.
The following reference strategies and measurements will address, and often
eliminate, the statistical and systematic uncertainties described in
\S~\ref{exec_uncertain}.

\addtolength{\partopsep}{-2mm}
\begin{itemize}
\item Blind, multiplexed searching.
\item SNe~Ia at $0.1 \le z \le 1.7$.
\item Spectrum for every supernova at maximum covering the rest
frame UV and Si~II 6150\AA\ feature.
\item Spectral time series of representative SN~Ia, with cross-wavelength
relative flux calibration.
\item Light curves sampled at frequent, standardized intervals starting
shortly after explosion and extending to 80 restframe days after explosion
to obtain a
light-curve-width- and extinction-corrected peak rest-frame $B$ brightness to 10\%.  This requires percent-level peak-magnitude determinations
in the supernova-frame optical passbands.
\item Multiple color measurements in a filter set consisting of
9 bands approximating
rest-frame $B$ at different redshifts.
\item Final images and spectra after the supernova
is no longer present to enable clean subtraction of 
host galaxy light. 
\end{itemize}
\addtolength{\partopsep}{2mm}

The quality of these measurements is as important as the 
time and wavelength coverage, so we require
control over $S/N$ for these photometry and spectroscopy 
measurements to give high statistical significance 
for supernovae over the entire range of redshifts.
We also require control over calibration for these photometry and spectroscopy 
measurements, by collecting monitoring data to measure 
cross-instrument and cross-wavelength calibration.

Note that to date no single SN~Ia has ever been observed with this
complete set of calibrated measurements, either from the ground or in space, and
only a handful have a data set that is comparably thorough.  With the
observing strategy described here, {\em every one} of $>2000$
followed SN~Ia will have this complete set of measurements.

The
requirements on spectroscopic exposure times and signal-to-noise are
described in detail in \citet{bk:03}.  Using an
exposure time of 8.5 hours for $z=1.7$ supernovae will provide
the desired spectral-parameter measurements
given the SNAP spectrograph parameters described in \S\ref{spec:sec}.
Exposure times for supernovae at lower redshifts down to $z \sim 1$
obey a scaling factor of $(1+z)^6$.  The exact exposure time necessary
will be refined as the library of
UV--optical supernova spectra continues to grow in coming years.

\begin{figure}[h]
\plotone{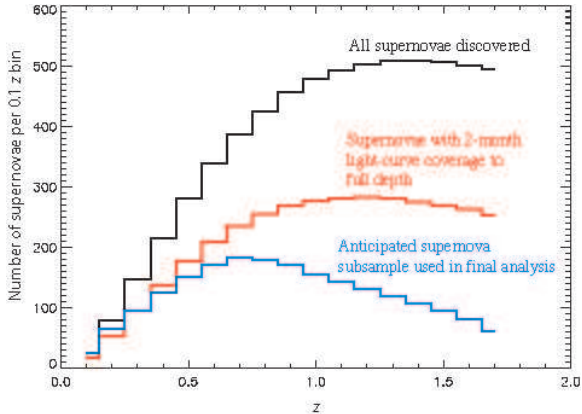}
\caption{Close to 6000 Type Ia supernovae
will explode in the SNAP field of view out to $z=1.7$ during
the course of the SNAP deep survey (upper black curve).
A subset of these will be observed in all filters to full
depth and early enough during the survey to be followed photometrically
for at least two months in the supernova frame (middle red curve).
Most of these objects will be observed spectroscopically.
A smaller subset of supernovae with homogeneous light curves and
spectra will be used in the final cosmology analysis; we anticipate
using a flat distribution in cosmic time (bottom blue curve).
\label{sndist:fig}
}
\end{figure}

The number of supernovae that SNAP is expected to discover is based
on the  supernova rate per unit volume of \citet{rate_02} assumed
independent of redshift, which when extrapolated
is roughly consistent with the
discoveries of the GOODS/ACS search and
Subaru supernova search.  This is a conservative assumption compared
to rates calculated incorporating
the cosmic star formation rate
\citep{madauetal:1998}.
SNAP will be able to spectroscopically observe almost all the
supernovae with deep and good temporal photometric coverage, $\sim
4000$ supernovae in the 32-month survey.
\citet{frieman:2003} find
that in a resource-limited survey with systematic uncertainties
taken into account, using
a relatively uniform distribution at high redshift is optimal
for providing cosmological parameter measurements.
As a result, SNAP will not provide spectroscopic followup
for a fraction of the highest redshift supernovae, which
are more plentiful than the lower redshift supernovae.  For the 
cosmology analysis, we expect to cull
the sample of spectroscopically observed supernovae to $\sim 2000$
events that satisfy strict homogeneity and extinction cuts
to yield a flat distribution
in cosmic time as shown
in Figure~\ref{sndist:fig} .

\subsubsection{Weak-Lensing Program}
The strengths that make SNAP excellent for supernova observations
apply to lensing as well; a wide-field imager in space with stable and
narrow point-spread-functions can provide large survey areas, accurate
shape measurements, and high galaxy angular-surface densities.  The
nine optical and near infrared filters are necessary for accurate
photometric redshift determinations.  The SNAP deep-survey fields in
which the supernovae are found will thus serve as a deep lensing
field.  The SNAP wide-field survey has been tailored to resolve the
shapes of $\sim 100$ galaxies per square arcminute. Over one year
$\sim300$ million galaxies will be resolved.  For source galaxies with
an intrinsic ellipticity distribution $\sigma_{\epsilon}=0.31$
\citep{rhodes:2000}, this will achieve a signal-to-noise of unity on
the shear measurement in a 1 square arcminute cell, over 1000 square
degrees.  The reference SNAP deep survey will exceed even this density
requirement by a factor of three
\citep{masseyetal:2003}. 

Current plans for the wide-field survey call for each field to be
observed for 1200 seconds in each optical filter, and twice as long in
each NIR filter.  Each pointing will be divided up into 6 individual
200 second exposures which are needed to eliminate cosmic rays and
cover the spatial gaps between the detectors.  We are engaged in
determining the systematic uncertainties intrinsic to the SNAP
weak-lensing program to find the optimal survey area and allotted time
for the wide-field survey; initial estimates indicate 300-1000 square
degree field (5--16 months) are appropriate for the wide-field survey.

The on-board storage space is matched to the telemetry and these two
constraints limit how many images can be stored and subsequently sent
to Earth.  If the telemetry constraints can be eased, either through a
higher download rate (and greater on-board storage) or through data
compression, the exposure times can be made shorter, thus increasing
the area of the weak lensing survey.

A full description of the
SNAP weak-lensing program can be found in a separate series of papers
\citep{rhodesetal:2003, masseyetal:2003,refregieretal:2003}.


\subsubsection{Panoramic Survey}
\label{panoramic:sec}
A extremely wide-field ``panoramic'' imaging survey is potentially
useful for weak lensing and other probes of dark energy, cosmology,
and astronomy in general.  SNAP will provide a unique means for
performing wide-field optical imaging from space.  We thus explore the
depth obtained from a three-year SNAP survey covering 7000--10000
square degrees.  We do not address specific scientific programs made
possible from this survey, considering it as a possible legacy dataset
that can serve the astronomical community at large.

\subsection{Orbit}
The reference mission calls for a halo orbit about the second
earth-sun Lagrange point (L2). The L2 point is about 1.5 million
kilometers from the earth along the sun-earth line and has a very
stable thermal environment. There are no eclipses and heat from the
earth is constant and negligible. There are ample communication contact
opportunities from the ground to SNAP. A ground station would
communicate from 2-4 hours per day with the spacecraft. This orbit is
highly advantageous from the standpoint of achieving passive detector
cooling; a three square meter radiator that operates at 130K provides
45 watts of gross cooling capacity.

Having the SNAP fields near the ecliptic caps places the Sun at nearly
right angles to the viewing direction throughout the year. In an L2
orbit, the Earth and Moon will always be nearly at right angles to the
viewing direction and on the sun side of the spacecraft. We utilize
this viewing geometry in several ways. First, the solar panels can be
rigidly body-mounted on the sunward side of the spacecraft, avoiding
the cost, failure modes, and flexure of deployed panels. Second, the
passive cooling radiator can be rigidly located in permanent shadow on
the antisunward side of the spacecraft. Third, the stray light
baffling can be optimized for a limited range of solar roll and
elevation angles, and for a limited range of Earth elevation angles.

We plan to have the spacecraft perform 90 degree roll maneuvers every
3 months during the mission, to keep up with the mean ecliptic
longitude of the Sun. The detector layout in the focal plane
has a 90 degree roll symmetry
that allows photometric scans with identical filter coverage from
season to season (see \S\ref{imager:sec}).

\subsection{Telescope}
\label{telescope:sec}
The requirements placed on the SNAP telescope derive directly from the
science goals and the mission constraints. The wavelength coverage is
determined by the need to observe a number of filter bands across the
visible and NIR wavelength range spanning over 0.35 $\mu$m to
1.7 $\mu$m, and to conduct low-resolution spectroscopy of each
supernova near maximum light. This requirement effectively rules out refracting
optical trains and drives the telescope design toward all-reflective
optics.

The light gathering power is set by the need to discover supernovae
early in their expansion phases and to permit photometry and low-resolution spectroscopy near maximum light. These requirements can be
met with a minimum aperture of about two meters. Note that for a fixed
signal-to-noise ratio ($S/N$)
the exposure time for an isolated point source is proportional 
to the telescope aperture to the fourth
power with background-limited noise and diffraction-limited
optics.

Image quality is also
a factor in determining $S/N$ because of the
effects of natural zodiacal light and detector noise.  We intend to
achieve angular resolution near the diffraction limit for wavelengths
longward of one micron. For a two-meter aperture and one-micron
wavelength the Airy disk size is 0.11'' FWHM. To match this
diffraction spot size to the CCD pixel size ($\sim 10$
$\mu$m) one must adopt an effective focal length of about 20
meters. This focal length is also matched in the NIR where
HgCdTe detectors
with 18--20 $\mu$m pixels will observe out to a photon wavelength of
17000 \AA.

A large field of view
is needed to obtain multiplexing advantage: a single exposure can contain tens
of different supernovae that are in important phases of their
light curves.  The SNAP reference mission
has a field of view of the order one square
degree, of which about 0.7 square degrees will be instrumented by
detector pixels.
The ratio of working field area to diffraction patch
area is about 800 million, comparable to the total number of detector
pixels.

An excellent optical quality and low-level of distortion must be
maintained over the working focal plane. The image quality of the
telescope is driven in part by the $S/N$ requirement, and also by the
potential systematic supernova spectrum contamination by unwanted
light from the supernova host galaxy. A system Strehl ratio of 0.90 at
one micron wavelength is baselined, which corresponds to an RMS wavefront
error (WFE) of 0.05$\mu$m.  At 0.633 $\mu$m the Strehl ratio is 0.77.
(The Strehl ratio is the peak
monochromatic image irradiance divided by the theoretical peak
irradiance for the ideal diffraction limited image.)

\subsubsection{Optical Configuration}

Prospective launch vehicles (Delta IV, ATLAS V, SeaLaunch) and payload
fairing dimensions impose limits on the telescope size and its
mass.  An overall payload length of about 6 meters and a payload
diameter of about 2.5 meters will accommodate the SNAP observatory.
Through a series of packaging exercises we have explored ways to fit
the maximum length stray light baffle into available launch fairings,
and find that with a short optical package, $\sim 3$ m in length, and
a tall outer baffle, the required stray light rejection can be
achieved (\S\ref{stray:sec}).

\begin{figure}[h]
\plotone{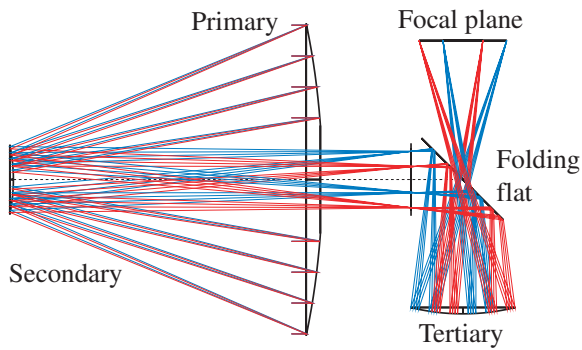}
\caption{Reference model SNAP optics layout. The entrance
pupil is defined by the 2-m primary mirror. The exit pupil is at the
folding mirror.
\label{optics:fig}}
\end{figure}

To accommodate dimensional limitations and wide-field optical quality,
the three mirror anastigmat described by
\citet{korsch77} is used.  A schematic view is shown in
Figure~\ref{optics:fig}.

Details of the reference SNAP optical configuration have been determined by an
iterative process involving exploring various alternative choices for
focal length, working field coverage, and packaging constraints
\citep{lampton:2002,lampton:2003}.
The optimized optical parameters are summarized in
Table~\ref{tel:tab}.  The overall length of the optical train is 3.3
meters. Compared with the 21.66 meter effective focal length, this
system has an effective telephoto advantage of about 6.5.  The mirrors
are pure conic sections of revolution having no polynomial terms. The
use of higher polynomial terms has not yet been explored. The location
of the vertex of each element is listed in a Cartesian (X,Z)
coordinate system whose origin is the vertex of the primary mirror and
Z is the optical axis.

\begin{deluxetable}{ccccccc}
\tablewidth{0pt}
\tablecolumns{7}
\tablecaption{Optical surfaces and locations\label{tel:tab}}
\tablehead{
Optic & Diameter & Central Hole & Curvature & Asphericity &X Location, &Z Location\\
 & (m) & (m) & (m$^{-1})$ & & (m) & (m)}
\startdata
Primary & 2.00 & 0.5 & -0.2037466 & -0.981128 & 0 & 0\\
Secondary & 0.45 & none & -0.9099607 & -1.847493 & 0 &-2.00\\
Folding flat & 0.80 $\times$ 0.48 & 0.20 $\times$ 0.12 & 0 & 0 & 0 & 0.91\\
Tertiary & 0.69 & none & -0.7112388 & -0.599000& -0.87 & 0.91\\
Focal plane & 0.577 & 0.258 & 0 & 0 & 0.9 & 0.91
\enddata
\end{deluxetable}

\subsubsection{Mechanical Configuration}
The launch environment imposes both stiffness and strength
requirements on the payload. Vehicle aeroelastic stability concerns
prescribe the needed payload stiffness in terms of minimum structural
frequencies in the axial ($\sim 25$ Hz) and lateral ($\sim 10$ Hz)
directions. The launch environment includes both quasi-steady and
random acceleration events that are combined to establish the peak
loads, or strength requirements for the payload.

\begin{figure}
\plotone{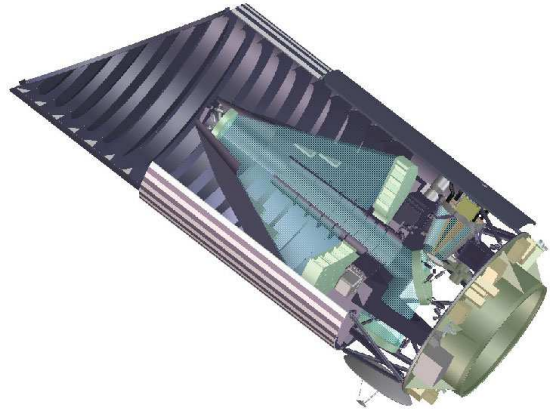}
\caption{Cutaway view of the reference SNAP design.
The entire telescope telescope attaches to the spacecraft structure by
means of bipods. The outer baffle, shown cut away, also attaches to
the spacecraft structure by means of its separate supporting struts. A
discardable door, shown open in light gray, protects the cleanness of
the optics until on-orbit commissioning begins. Solar panels are
fixed, not deployed.\label{cut:fig}}
\end{figure}

\begin{figure}
\plotone{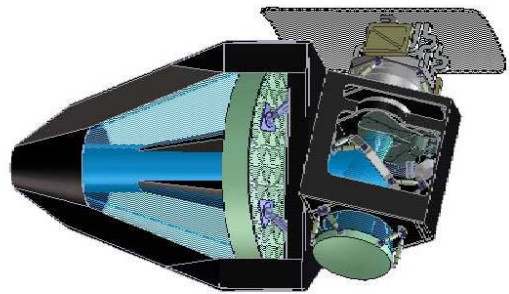}
\caption{Telescope metering structure (carbon fiber, shown in dark gray) provides precision control of optical element spacings and orientations. Forward of the primary mirror, the secondary is supported on adjusters within the secondary baffle. Aft of the primary mirror, the tertiary metering structure supports the folding flat, the tertiary, and the focal plane instrumentation. The passive radiator at top is thermally but not structurally linked to the focal plane instrumentation.\label{meter:fig}}
\end{figure}

For a space mission it is vital to create a mechanical configuration
which provides an extremely stable metering structure that maintains
the optical element alignment during ground testing, launch, and orbit
operations. The concept adopted for SNAP is to create three structural
components that will be brought together during spacecraft/payload
integration: a stiff low-precision outer baffle cylinder carrying the
exterior solar panels and extensive thermal insulation; a stiff
low-precision spacecraft bus structure that carries antennas,
batteries, and other major spacecraft support components; and a stiff
high-precision telescope structure comprising carbon-fiber metering
elements, the kinematically-mounted mirrors, the instrumentation
suite, and its own thermal control system. Figure~\ref{cut:fig} shows
the overall payload and spacecraft layout, while
Figure~\ref{meter:fig} shows details of the secondary and tertiary
metering structures.

\subsubsection{Materials}
Space-proven optical mirror technology is largely based on two
approaches: open-back Schott Zerodur glass ceramic composite material
and Corning ultra-low expansion ULE glass honeycomb structure. For
SNAP either technology has sufficiently low coefficient of thermal
expansion and sufficiently well proven manufacturing
techniques. Studies are underway exploring the detailed fabrication
and test flows using each process.

The metering structure will utilize a low-CTE carbon-fiber
construction. In particular, the secondary support tripod will have to
maintain the primary to secondary spacing accurate to a few
microns. This tripod, and the other major metering components, will
certainly require a dedicated active thermal control system
with passive cooling. We
anticipate the need for five-axis motorized adjustment for the
secondary mirror during ground integration, on-orbit observatory
commissioning, and occasionally during science operations. For this
reason we plan to include a hexapod or other multi-axis positioner
into the secondary-mirror support structure.

The single highest priority  bearing on the choice of
mirror coating is the system throughput at the longest wavelengths
where rest-frame optical photons from the  highest-redshift supernova
are the most distant and photons are the most
precious.  A secondary consideration is to establish a low thermal
emissivity for the mirrors so that operating them at approximately
290K will not seriously impair our astronomical
sensitivity in the NIR bands.  The most common coating for
astronomical mirrors at visible wavelengths is SiO overcoated
aluminum.  It offers outstanding durability and unmatched reflectance
throughout the visible band, 0.4 to 0.7 $\mu$m.  A less common choice
is protected silver, which is less efficient in the blue but more
efficient in the red and NIR.  The SNAP optical system
has four reflections so its throughput varies as the
fourth power of the mirror coating reflectivity.  We have assumed here
the use of protected silver rather than protected aluminum owing to
its substantially higher total system throughput at the photon-starved
 wavelengths.

\subsubsection{Geometric-Optics Performance}
In the reference SNAP optical configuration, the center of the focal plane is
severely vignetted by the necessary hole in the
folding flat which lies near the Cassegrain
focus.  In optimizing the optical performance of the system,
the useful imaging surface on the focal plane is taken to have
an annular shape.

\begin{figure}[h]
\plotone{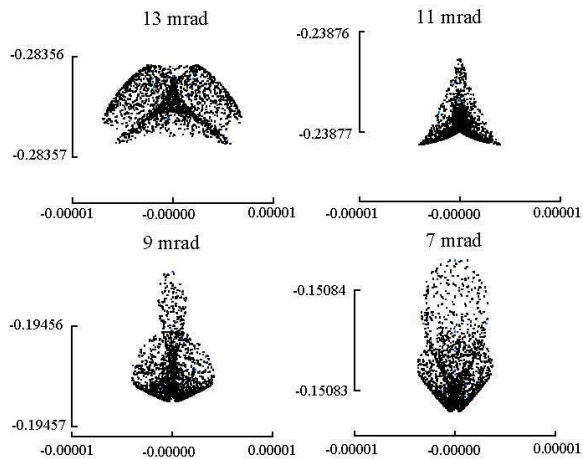}
\caption{Ray trace spot diagrams. Upper left: 13 mrad off axis; upper right 11 mrad; lower left 9 mrad; lower right 7 mrad. Tick marks are spaced 10 $\mu$m in the focal plane.\label{spot:fig}}
\end{figure}

The optical performance of our reference telescope is
fundamentally limited by aberrations and manufacturing errors at
short wavelengths, and by diffraction at long
wavelengths. Accordingly, our expected performance figures divide into
two areas: geometrical ray traces that quantify the aberrations
and pupil diffraction studies. We summarize the key performance
items in Table~\ref{opt:tab} and Figure~\ref{spot:fig}.

\begin{deluxetable}{cc}
\tablewidth{0pt}
\tablecaption{Reference optical performance\label{opt:tab}}
\tablehead{Parameter&Value/Performance}
\startdata
Focal Length & 21.66 meters\\
Aperture & 2.0 meters\\
Final focal ratio & f/10.83\\
Field & Annular, 6 to 13 mrad; 1.37 sq deg\\
RMS geometric blur & 2.8 $\mu$m, average 1 dimension\\
Central obstruction & 16\% area when fully baffled\\
Vane obstruction & 4\% area, tripod\\
\enddata
\end{deluxetable}

From spot diagrams at various off-axis angles, we have compiled the
statistics on the mean radial centroid of ray hits in the focal plane,
and the second moments of the spot distributions. These are listed in
Table~\ref{moment:tab} in the form of the two orthogonal RMS
breadths (radial RMS and tangential RMS) in columns 2 and 3. The
mean breadth is
listed in column 4 below as microns on the focal plane,
and in column 5 as an angular size on the sky in milliarcseconds.  At
the inner and outer radii of the image annulus, the FWHM becomes as
large as 60 milliarcsec, although in the midrange of the annulus it
is smaller.

\begin{deluxetable}{cccccc}
\tablewidth{0pt}
\tablecaption{Off-axis angle, image moment, and radial distortion\label{moment:tab}}
\tablehead{
Off Axis &Radial RMS, &Tangential RMS& Average RMS & Average RMS 
&Radial Distortion\\
$\sin{\theta}$ & ($\mu$m)&($\mu$m) & ($\mu$m) & (milliarcsec) &  ($\mu$m)}
\startdata
 0.006  &3.32 &1.60 &2.62 & 24.9   &-838 \\
0.007&  3.33 &1.60 &2.62 &24.9  &-782 \\
 0.008  &3.18 &1.59 &2.53 &24.1  &-630 \\
0.009 &2.83 &1.51 &2.28 &21.7  &-373\\
 0.010  &2.28& 1.37& 1.89 &18.0  &0\\
0.011 & 1.57 &1.35 &1.47 &14.0  &509 \\
 0.012  &1.18 &1.89 &1.58 &15.0  &1165\\
 0.013&  2.09& 3.23 &2.73 &26.0 &1983\\\hline
&& AVERAGE= &2.22& 21.0 \\
\enddata
\end{deluxetable}

Distortion is another fundamental optical aberration, but it does not
impact the $S/N$ nor does it directly impact the detection of
supernovae unlike the other Seidel aberrations distortion. It does
however cause the loci of scanned field objects to depart from
parallel tracks in the focal plane, and does complicate the weak
lensing science.  In our reference design, we have disregarded
distortion as a driver, in order to use all available design variables
to maximize the working field of view and minimize the net geometrical
blur. It is nonetheless important to explore the resulting distortion
quantitatively. The telescope mirror assembly (TMA) distortion is axisymmetric owing to the
symmetry of the unfolded (powered) optical train, and in polar
coordinates any off-axis angle maps onto a single focal plane radius
independent of azimuth angle. The distortion is therefore purely
radial. Column 6 of Table~\ref{moment:tab} lists the radial distance
of an off-axis field point as a function of the sine of the off-axis
angle, and the departure from proportionality to the sine of that
angle.  The TMA distortion is of the pincushion type, having increased
magnification toward the extremity of the field. Compared to a linear
mapping of $\sin{\theta}$ onto focal plane radius, the distortion
amounts to about two percent.

\subsubsection{Pupil Diffraction}
For a star at infinity and a telescope focused at infinity, the pupil
diffraction pattern is computed using the Fraunhofer formalism.  The
focal plane irradiance is simply the square of the modulus of the
two-dimensional Fourier transform of the pupil. We have computed this
irradiance function and diffracted light background for a variety of
prospective metering structures.

\subsubsection{Stray Light}
\label{stray:sec}
\begin{figure}
\plotone{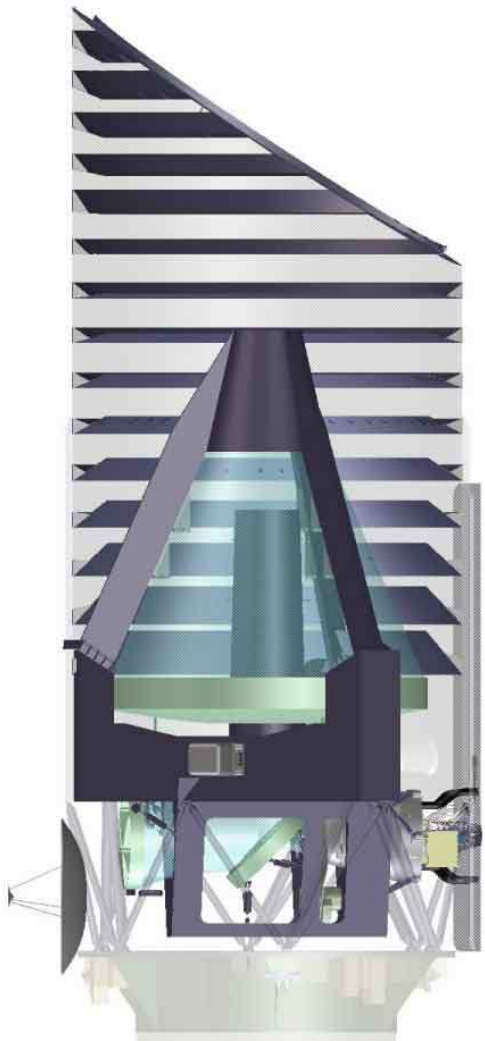}
\caption{Schematic treatment of the outer baffle interior vane arrangement. Sunlight is incident from the left, where the height of the baffle and its angled forward edge maintains the baffle interior in darkness.\label{baffle:fig}}
\end{figure}

A comprehensive stray light control plan has been developed for
SNAP. The goal is to keep all stray light sources far below the
natural zodiacal irradiance level as seen at the focal plane. The
primary concern is sunlight scattered past the forward edge of the
outer light baffle. This requires a minimum of two successive forward
edges, since the light diffracted past a single edge would exceed the
allowable irradiance at the primary mirror, assuming typical mirror
scattering values.  Figure~\ref{baffle:fig} shows
our treatment of the outer baffle interior vane arrangement. 
 The mission reference L2 orbit simplifies the stray
light design by eliminating the possibility of sun and full Earth light
entering the main baffle from opposite directions. Analysis efforts
are underway to quantify the effects of both out of band light
rejected by the focal plane filters, as well as ghosting and internal
reflections within the filters.  A systems engineering approach is
being pursued, along with development of a detailed stray light budget
for both solar and stellar contamination sources.

\subsubsection{Tolerances}
The departure of any surface from its nominal mathematical conic
section or the misplacement or misorientation of any of the surfaces
causes a wavefront error and a degraded image quality. One measure of
this degradation is the telescope's Strehl ratio. The Strehl ratio can
be converted into RMS wavefront error (RMS WFE) through Marechal's
relation. To achieve a system Strehl ratio of 0.77 at 0.633 $\mu$m
wavelength, the total WFE must not exceed 0.05 nm RMS. This allowed
WFE will be apportioned into individual contributions as part of the
detailed telescope design.

A tolerance budget has been developed based on a group of exploratory
studies of the sensitivity of the geometrical spot size to variations
in element curvature, shape, location, and orientation. These
calculations show that by far the single most critical parameters are
the primary mirror curvature and the spacing between primary and
secondary mirrors expected given the fast ($f/1.2$) primary mirror. A
two-micron displacement of the secondary piston, or a two-micron
displacement in the virtual image created by the primary mirror is
found to increase the RMS geometrical blur by about 3 $\mu$m.
Similarly, a 15-$\mu$m lateral displacement or a 15-microradian tilt
of the secondary mirror causes a 3-$\mu$m growth in the
RMS geometrical blur.

The reference model SNAP telescope includes on-orbit mechanical adjustments
that permit the relocation and reorientation of the secondary mirror,
and possibly the tertiary mirror as well, to optimize image
quality. By means of these adjustments we anticipate accommodating
small shifts in any of the optical elements locations and
orientations, allowing for correction of geometrical blur.

\subsection{Imager}
\label{imager:sec}

The wide field of view of the SNAP imager allows simultaneous batch
discovery and building of supernova rest-frame optical
light curves, and over the mission lifetime will yield
several thousands of supernovae at $0.1<z<1.7$
with the proposed photometric
accuracy.  More distant, less precisely measured supernovae will also
be available in our data set.  The multiplexing advantage with this
large field reduces by more than an order of magnitude the total exposure times
necessary to follow all the supernovae.  Figure~\ref{syslc:fig} shows
critical points on the light curve and the desired measurement
accuracy that the SNAP imager must furnish. We note that the stated
$S/N$ need not be achieved with a single measurement but can be
synthesized from multiple measurements, taking advantage of the
substantial time dilation for high-redshift supernovae.

Weak lensing by large-scale structure is a statistical study that
requires the accurate measurement of the shapes of many  galaxies.  In
order to efficiently survey a large area in a finite time, a large field
of view covered by detectors with  high quantum efficiency is necessary.
Given a fixed number of pixels in the focal plane, there are competing
desires to  have better resolution, which requires smaller pixel scales
and wider sky coverage, which requires larger pixel scales.   The
theoretical optimum of this trade-off requires that the pixel size be roughly
matched to the size of the PSF \citep{bernstein:etc};
further
plate scale and pixel size studies using both simulated and real data
are currently
in progress for optimizing photometric and shape-measurement
accuracy while maintaining a large field of view.
For the case of  the state-of-the-art
SNAP CCDs in which the PSF size is dominated by diffusion 
in the silicon at $\sim8000$
\AA, the theoretical optimum requires a pixel  size of roughly 0.1 arcsecond.  (Diffusion
would be worse in a conventional CCD that has a field-free region.)
In order to make precise, bias-free photometric redshift measurements,
wide spectral coverage for the optical to near infrared is
needed. Such photo-z measurements are crucial to interpreting the
cosmological implications of the wide-field survey.  The
aforementioned capabilities are necessary to measure the shapes of a
significant number of faint galaxies beyond $z=1$ in order to study
the evolution of structure.  Crucially important systematic errors in
shape measurements arising from unstable anisotropic PSFs and
systematic errors/biases in photometric redshift determination are
significantly reduced in a space-based imager such as SNAP when
compared to ground-based facilities.

The SNAP imager addresses the above requirements using two detector
technologies to efficiently cover the wavelength range of 3500 \AA\ to
17000 \AA.  The visible region (3500 \AA\ to 10000 \AA) is measured with
Lawrence Berkeley National Laboratory new-technology
$p$-channel high-resistivity CCDs
\citep{holland03,stover00,groom00} which have high ($\sim$80\%) quantum
efficiency for wavelengths between 3500 and 10000~\AA. Extensive
radiation testing shows that these CCDs will suffer little or no
performance degradation over the lifetime of SNAP
\citep{traps2002,protons2002}.  In the
reference design a pixel size of
10.5 $\mu$m has been matched to the telescope focal length and
diffraction limit at
10000 \AA\ of 0.1 arcsec. Improved shape measurements
and robust PSF photometry
come from a dithered observation strategy.
Undersampling, dithering, cosmic ray hits, and
many other effects are included in the exposure time calculator
developed by \citet{bernstein:etc}.  

The NIR range (9000 \AA\ to 17000 \AA) is measured with commercially
available HgCdTe arrays. Current large area (2048 $\times$ 2048 pixel)
HgCdTe detector devices have a pixel sizes in the range of 18 - 20
$\mu$m.  The telescope optics are designed to give an angular pixel
size of 0.17 arcsec, matched to the telescope focal length and diffraction limit
at 17000 \AA. The SNAP reference performance specifications require low
read noise and dark current ($<$ 6-8 e$^-$ for multiple reads and $<$ 0.1
e$^-$/sec/pix, respectively), high quantum efficiency ($>$60\%), and
uniform pixel response.  Reference imager detector specifications are given in
Table~\ref{ref:tab}.  The evolution of the SNAP
focal-plane design is described in \citet{lamptonfp:2002}.

\begin{deluxetable}{ccc}
\tablewidth{0pt}
\tablecaption{Reference mission specifications for the imager and its sensors\label{ref:tab}}
\tablehead{
Parameter & Visible & NIR }
\startdata
Field of View (deg$^2$)& 0.34 & 0.34\\
Plate scale (arcsec) & 0.10 & 0.17\\
Wavelength (\AA) & 3500--10000 & 9000 -- 17000 \\
$\left< \mbox{Quantum efficiency} \right>$ (\%) & 80 & 60\\
Read noise (multiple reads) (e$^-$/pixel) & 4 & 6--8 (multiple read)\\
Dark current (e$^-$/s/pixel) & 0.002 & 0.1 \\
Diffusion ($\mu$m) & 4 & 5\\
Number of fixed filters & 6 &3\\
\enddata
\end{deluxetable}

The minimum filter set required is primarily determined by the
precision needed for the cross-filter $K$-correction, that is
necessary when reconstructing the primary restframe band ($B$-band in
the reference design) light from a set of laboratory-frame filter
measurements. Six visible filters and three NIR filters are sufficient
if they are derived from a B-band filter with logarithmic $(1+z)$
scaling of their wavelength centers and widths. The reference set that
we consider here consists of nine redshifted $B$ filters
logarithmically distributed in wavelength with effective wavelengths
at $4400 \times 1.15^n$\AA\ for $n \in \{0,1,...,8\}$.  Further
optimization of the filter set for supernova science and
photometric-redshifts is in progress.  In particular, the ultimate
SNAP ``$B$'' filter will not necessarily be a Johnson $B$ although it
is expected to have a similar central wavelength and width.

\begin{figure}
\plotone{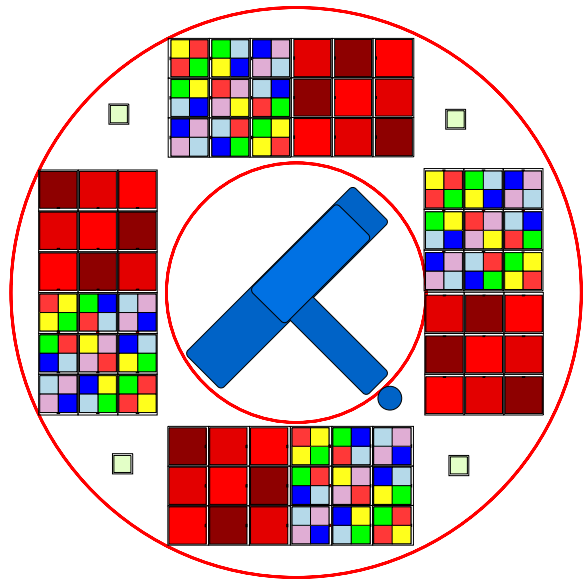}
\caption{The SNAP focal plane working concept. The two-axis symmetry of the imager filters allows any 90$^\circ$ rotation to scan a fixed strip of the sky and measure all objects in all nine filter types. 
The imager covers 0.7 square degrees.  Underlying the filters, there
are 36 2k x 2k HgCdTe NIR devices and 36 3.5k x 3.5k CCDs on a 140K
cooled focal plane.  The six CCD filter types and three NIR filters
are arranged so that vertical or horizontal scans of the array through
an observation field will measure all objects in all filters.  The
false colors indicate filters with the same bandpass.  The central
rectangle and solid circle are the spectrograph body and its light
access port, respectively. The spectrum of a supernova is taken by
placing the star in the spectrograph port by steering the
satellite. The four small, isolated squares are the star guider
CCDs. The inner and outer radii are 129 and 284 mm,
respectively.\label{fplane:fig}}
\end{figure}

A fixed-filter design is preferred over
a large filter wheel which would be a single point failure risk.
Figure~\ref{fplane:fig} shows an array
of visible and NIR filters in the SNAP focal plane.
To enhance the amount of NIR light that is
integrated, the individual NIR filters have twice the area in the focal
plane of the
individual visible filters.
The constraint that the satellite be rotated in 90$^\circ$ increments
requires the filter pattern to remain symmetric with respect to two
orthogonal axes. Consider the filter arrays in Figure~\ref{fplane:fig}. Note that the arrays can be scanned though an observation field
left-to-right, right-to-left, top-to-bottom, and bottom-to-top, and
that a given star will be measured with each filter bandpass but not
necessarily the same physical filter. Note that any 90$^\circ$ rotation of
the filter array can still measure the star field in all filter types.

As shown in
Figure~\ref{fplane:fig},
underlying each NIR filter is one 2k
$\times$ 2k, 18 $\mu$m HgCdTe device, 36 in total. Underlying each
2$\times$2 array of visible filters is one 3.5k $\times$ 3.5k, 10.5
$\mu$m CCD, with 36 devices in all. The total number of pixels is
$\sim$600 million.

The SNAP focal plane will be passively cooled to operate at 140K, a
habitable temperature for the CCDs and sufficiently cold to achieve
the noise requirements from the 1.7$\mu$m-cutoff HgCdTe devices.
Short flex cables penetrate the focal plane to bring the signals to
the electronics located on the backside. The present reference design
envisions an ASIC-based readout that would also operate at 140K; this
avoids routing low-level analog signals long distances and reduces the
size of the cable plant between the cold focal plane and the warm data
acquisition electronics located to the side.

A four-channel custom
chip will read out each CCD \citep{walder:2003}. It consists of a
single-ended to differential converter followed by a correlated double
sampler and a novel multi-slope integrator. This circuit is designed
to operate at room temperature for test purposes and at 140 K. The
readout speed is 100 kHz. The 16-bit dynamic range is covered using 3
gains each with a 12 bit signal to noise ratio. A 12-bit pipeline ADC
digitizes the analog data.

In the deep survey,
multi-band exposures of a point of sky will be achieved by
shift-and-stare observations in which the pointings are shifted by the
2.9 arcmin width of the optical filters.  Each pointing will consist of four
300-second exposures; the multiple exposures are for cosmic-ray
rejection and dithering of our undersampled pixels.
The fixed filters prevent independent tuning of exposure times for
different filters.
Within a scan,
over one hundred pointings are required to cover the 7.5 square
degrees in all filters to the desired depth.  A
scan of the north (south) field will be repeated every four days for
16 months for a total 120 scans.

In the wide (and panoramic) surveys, long adjacent strips will be
observed using the shift and stare method.  Six rather than four
dithers will be made per pointing so as to fill the gaps between the
detectors and provide robust galaxy-shape measurements.  Observed
patches of sky will not be revisited.  The exact shapes of the fields
are under review.

\subsection{Spectrograph}
\label{spec:sec}

The spectrometer is used to make a positive identification of Type Ia
supernovae by observing the characteristic SiII feature at 6150\AA,
measure the features associated with supernova heterogeneity,
determine the redshift of the underlying host galaxies and supernovae, build a
template spectral library of SNe Ia, and link standard stars with
the SNAP photometric system.

The specific signature of Type Ia supernovae is the SiII line at
$\lambda=6150$\AA\ (rest frame).  In models, the strongest sensitivity
to the metallicity in the progenitor system lies in the rest-frame UV
band.  The need for both supernova-frame UV and Si features defines a
broad wavelength range, 4000 - 17000 \AA\ that must be covered by the
instrument to observe $z \le 1.7$ supernovae. 
The
broad features of the supernova spectra and the non-negligible detector
noise contribution for the faintest objects make a low-resolution
spectrograph optimal: studies \citep{bk:03} find an optimal resolving
power $\lambda/\delta\lambda \sim 100$ at FWHM and 1 pixel per FWHM
sampling, with constant resolving power in the 0.6-1.7$\mu$m range.
 Other types of supernovae, such as
II or Ib, have lines of H or He in the same wavelength range with
similar equivalent widths, allowing
supernova classification of all possible candidates with
one spectrograph.

The silicon feature and other features in the spectrum
of a typical Type Ia supernovae, shown in Figure~\ref{sysspec:fig},
are to be carefully observed.
Their  characteristics (position, width,
height, etc.) are related to the supernova peak magnitude through
physical parameters such as temperature, velocity and progenitor
metallicity.

The field of view must include the underlying galaxy in order to
determine its spectrum during the same exposure. This is necessary for
subtraction of the host spectrum from the spectrum in the supernovae
region, and for an accurate determination of supernova redshift. 
To avoid single-point failure in the spectrograph we desire
two detectors in each focal plane.  The detectors cover
3'' $\times$ 6'' split into 40 slices that
can easily cover
the mean size of galaxies at redshift 1-2.  The large field
of view allows the pointing requirements to be relaxed.
 The main spectrograph specifications are
summarized in Table~\ref{spec:tab}.
See \citet{ealet02} for details on the design process.

\begin{deluxetable}{ccc}
\tablewidth{0pt}
\tablecaption{Reference mission spectrograph specifications\label{spec:tab}}
\tablehead{
Property &Visible &NIR}
\startdata
Wavelength coverage ($\mu$m) &0.35-0.98 &0.98-1.70\\
 Field of view &3.0'' $\times$ 6.0'' &3.0'' $\times$ 6.0''\\
 Spatial resolution element (arcsec) &0.15& 0.15\\
 Number of slices& 40 &40 \\
Spectral resolution, $\lambda/\delta\lambda$ & 100 &100\\
\enddata
\end{deluxetable}

\subsubsection{Integral Field Unit}
The requirement for simultaneous acquisition of supernova and host spectra,
the ability to subtract off the host-galaxy spectrum after the
supernova has faded,
spectrophotometry for calibration purposes, and the high object
acquisition precision that would be needed for a traditional long slit
spectrograph, make a 3D spectrograph an attractive option. A 3D
spectrograph reconstructs the data cube consisting of the two spatial
directions X and Y plus the wavelength direction. For each spatial pixel, the spectrum is
reconstructed.  Two principal techniques are indicated for
3D spectroscopy: first, the use of a Fourier Transform Spectrometer
(FTS), and second, the use of integral field spectroscopy.

%

Trade studies indicate that  integral-field
spectroscopy using the image slicer
technique is preferred for SNAP. This technique, developed since 1938
in order to minimize slit losses, is very powerful
\citep{bowen38,content:1997,prietoetal:2002}. The new generation of image slicers improves the
efficiency and the compactness of the system and appears to be a very
well adapted solution for the SNAP mission.

\begin{figure}
\plotone{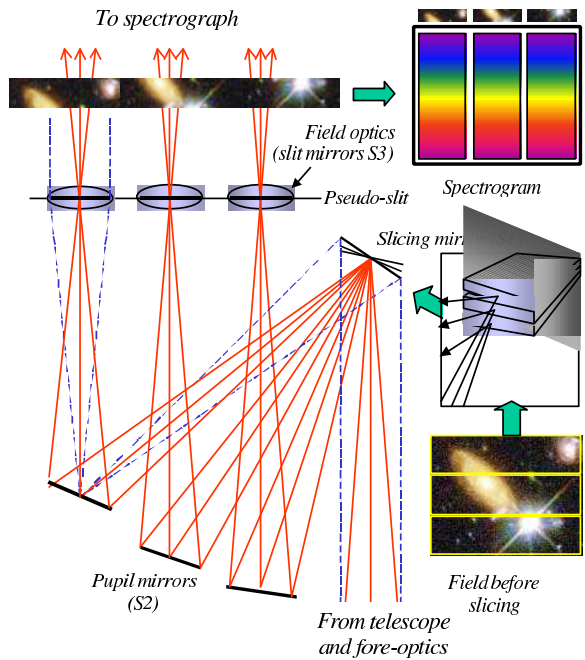}
\caption{Principle of the image slicer.
The field of view is sliced along $N$ (here $N=3$)
strips on a slicing mirror. Each  slice
re-images the telescope pupil onto  a line of
``pupil mirrors'' which reimages the field strip along a ``pseudo-slit''.
The pseudo-slit is
placed in the entrance plane of the spectrograph, acting as the
entrance slit (courtesy J. Allington-Smith, Durham U.).\label{slicer:fig}}
\end{figure}

Figure~\ref{slicer:fig} shows the principle of this
technique. The field of view is sliced along $N$ (in the drawing $N=3$,
for SNAP $N=40$) strips on a slicing mirror, consisting
of a stack of $N$ plates where the
active surface is on an edge. Each of the $N$ slices
re-images the telescope pupil, creating $N$ telescope pupil images in
the pupil plane. Thanks to a tilt adapted to each individual slice,
the $N$ pupil images lie along a line. In the pupil plane, a line of
``pupil mirrors'' is arranged. Each pupil mirror is placed on a pupil
image and reimages the field strip. These images are arranged along
a line and form a ``pseudo-slit''.  At this stage, therefore, there is an
image of each of the $N$ strips of the field of view. The pseudo-slit is
placed in the entrance plane of the spectrograph, acting as the
entrance slit.
A last line of mirrors is placed on the pseudo-slit. This line adapts
the output pupil of the slicer into the input pupil of the
spectrograph.

\subsubsection{Reference Spectrograph Instrument Concept}
\begin{figure}[h]
\plotone{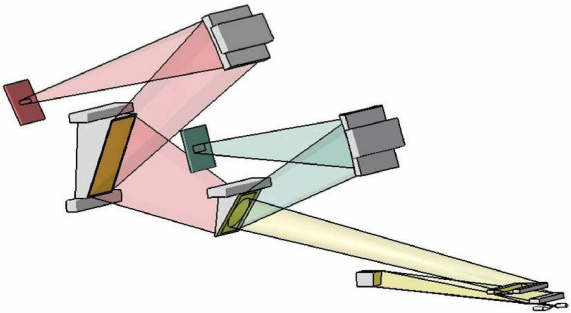}
\caption{
A schematic spectrograph optical design. The beam is going out from
the slicer (on the bottom right) to a prism disperser back faced
coated by a dichroic. The visible light (blue path) is reflected and
the IR light (in red) continue to a second prism used to reach the
required spectral dispersion. The two beams are therefore focused
on two detectors.  The dimensions of the spectrograph are approximately
$400 \times 80 \times 100$ mm.
\label{sblock:fig}}
\end{figure}

The spectrograph components are summarized in the block diagram
shown in Figure~\ref{sblock:fig} with the principal components
described below.

\subsubsubsection{Relay Optics}

This unit is the interface between the telescope beam and the
instrument. Some telescope
aberrations can be corrected within this optical system. A simple
three-mirror configuration should be sufficient to allow picking
off the beam at a point most convenient for the spectrometer.

\subsubsubsection{Slicer Unit}
The slicer unit acts as a field reformatting system. As described
above, the principle is to slice a 2D field of view into long strips
and optically align all the strips to form a long spectrograph
entrance slit. The slicing mirror is comprised of a stack of
slicers. Each slicer has an optically active spherical surface on one
edge. A line of pupil mirrors does the
reformatting. Each pupil mirror sends the beam to a slit mirror, which
adapts the pupil to the entrance of the spectrograph.

The long thin active surface of each individual slicer will produce a
large diffraction effect. In order to minimize flux losses to a few
percent, the spectrograph entrance pupil must be oversized. A combined
theoretical and experimental approach is underway at Laboratoire
d'Astrophysique de Marseille to define the optimum entrance pupil for
a JWST application (in the infrared bands 1-5 $\mu$m). This will
 be directly adapted for the SNAP concept.

The reference requirements on the slicer unit are an accuracy of $\le
\lambda/10$ RMS on the optical surfaces and a surface roughness of
$\le 50$ \AA\ RMS. Existing prototypes fully meet these specifications.

\subsubsubsection{Optical Bench}
Thanks to the moderate beam aperture and field of view, the
spectrograph optics will be straightforward. The reference is a
classical dichroic spectrograph: one collimator mirror, one prism with
a dichroic crystal, and two camera mirrors are required. Using
spherical shapes for all the mirrors would provide an adequately sharp
image, but using aspheric mirrors will make it possible to have a very
compact spectrograph. The prism solution is well matched to the
requirement of a constant resolution over the whole wavelength range. The
dichroic crystal allows the coverage of two channels simultaneously: one for
the visible (e.g.. 0.35--0.98 $\mu$m) and one for the infrared (0.98--1.70
$\mu$m).

\subsubsubsection{Detectors}
In the visible, the main goals are high quantum efficiency and very
low noise. Given concerns over degradation due to radiation exposure
and the poor performance of conventional thinned CCDs in the red part
of the visible, the fully-depleted LBNL CCDs are the detectors
of choice. Thinned, back-side
illuminated, deep depletion,
low-noise conventional CCDs of 1024 $\times$ 1024 pixels
are an alternative option. In the NIR, some factors constrain the
detector technologies. The operating temperature will be dictated by
the detector dark current. HgCdTe arrays with cutoff wavelength of 1.7
$\mu$m are currently under consideration which allow operation in the
130-140 K range. A detailed list of the performance specifications for
the detectors is provided in Table~\ref{specdec:tab}. To achieve the
listed performance in read noise and dark current, a multiple sampling
technique is required. To optimize exposure time, the impact of the
rate of cosmic rays on the spectrum quality is under study.

\begin{deluxetable}{ccc}
\tablewidth{0pt}
\tablecaption{Reference mission spectrograph detector specifications\label{specdec:tab}}
\tablehead{
& Visible &NIR}
\startdata
 Detector size &1k $\times$  1k &1k $\times$ 1k \\
Pixel size ($\mu$m) &10-20 &18-20 \\
Detector temperature(K) &140 &130-140\\
 $\left<\mbox{QE}\right>$(\%) &$>$80 &$>$60 \\
Read noise(e$^-$/pixel)& 2 &5 (multiple read)\\
Dark current(e$^-$/pixel/s)& 0.001&0.02\\
\enddata
\end{deluxetable}

\subsubsection{Efficiency Estimate}
Simulations of the efficiency of the instrument show a cumulative
efficiency of the instrument from the relay optic and
including the detector at a
level better than 50\% in the visible and 40\% in the infrared.
This excellent performance is due to high slicer efficiency and is
based on a conservative
value for the efficiency of individual mirrors (98\%), much
lower than is expected from a silver coating at these wavelengths. The
principal losses are from the prism (also conservatively estimated at
80\%), and from the detector.

\subsection{Telemetry}
During the reference deep survey,
the observation time of SNAP is partitioned roughly 60\% to preprogrammed,
stepped photometry
and 40\% to targeted spectroscopy with parallel photometry. The former
is divided into a sequence of 300s exposures followed by 30s of sensor
readout.  The latter is comprised of exposures varying from a few
seconds to a thousand seconds. For the longer exposure times, the
spectrograph NIR detector is continuously read in up-the-ramp mode to
allow cosmic-ray rejection.  Imaging is possible during spectroscopic
exposures; the imager is read as in photometry mode except that the
exposure times vary, determined by the time to achieve a required
$S/N$ in the spectrograph for a particular supernova.  For each imaging
exposure, the CCD devices produce a single frame of correlated double
sampled data while the NIR devices produce two frames (each containing
the average of multiple reads), one for post reset values and the
other for post integration values so that the digital correlated
double sampling can be done on the ground.

The operating concept for processing the raw data is to perform only
lossless compression of the frames in the satellite. A
factor of two is easily done; a greater compression factor is
probably achievable. The primary motivations are to be able to look
retrospectively into the data to better determine a supernova explosion time
and to be able to co-add many months of data for weak lensing science to
extract weak signals from the per frame noise. With this approach we
avoid developing flight software for automated acquisition and
processing of reference, dark and flat frames and applying them
irrecoverably to data frames. The consequence is that in L2 SNAP will require
about 133 Gb of flight data storage for a 24 hour contact period
through the Deep Space Network capacity.  We anticipate that DSN
will have the ability to receive 300 Mbit s$^{-1}$ Ka-band data
by the time SNAP launches.

The current reference weak lensing survey \citep{rhodesetal:2003} calls for a
1-year 1000 square degree survey consisting entirely of imaging. At
L2, 6 dithered exposures per optical filter would be taken for a total
of 1900 s. Such a survey would require a daily four hour download of
data and would use only $\sim150$ GB of on-board storage.  A wider
survey could be taken if more time were given to the survey or if the
exposure time per filter were lowered.  The latter option would
require that the data download time be accordingly raised to
compensate for more exposures or that the on-board compression of the
data be increased beyond the factor of two that can be achieved
through lossless compression.

\section{Calibration Program}
\label{calib:sec}

An important feature of the supernova-cosmology analysis is its
dependence on the {\it relative} brightness of Type Ia
supernovae. Cosmological and dark-energy parameters are determined
from the shape, and not the absolute normalization, of the Hubble
brightness-redshift relationship.  In order to constrain the possible
effects of population evolution, SNAP will use homogeneous subsets of
SN Ia.  Identification of these SN Ia subsets depends in part on the
brightness and color evolution of supernovae. Each supernova redshift
$z$ is plotted against its restframe B band flux which depends on the
K-correction and the extinction correction, which in turn depend on
the color of the supernovae.  As a result, the analysis relies on the
{\it color}, where we use this term to mean the flux ratios of
calibrated passbands.  The SNAP calibration program must thus
precisely control the absolute {\it color} calibration, not
necessarily the absolute flux calibration, in the $0.35 - 1.7$ $\mu$m
observing window.

The SNAP calibration program will provide fluxes in physical units
associated with the same restframe passband for a homogeneous set of
supernovae distributed out to $z=1.7$.  Calibration provides the
transformation of the measured signal in a given passband into a flux
associated with a calibrated passband.  Calibration affects
flux-correction for extinction due to the Milky Way, the SN host
galaxy, and the intergalactic medium, and also the K-corrections which
provide the transformation between fluxes in the observed and
restframe passbands.

In the SNAP cosmological-parameter determination, calibration enters
early in the analysis and consequently the calibration uncertainty
propagates non-trivially throughout.  Depending on its nature and
size, the calibration uncertainty can contribute significantly to the
$w_0-w'$ error budget. The propagation of these uncertainties has been
explored in \citet{klmm:03} for a simplified blackbody calibration
model.  Realistic calibration uncertainty will be more complex and the
SNAP calibration methodology is being crafted to ensure that
calibration does not impede achievement of the science goals.

The SNAP reference calibration concept consists of several programs
which are subject to ongoing work examining modifications, extensions,
and alternative strategies.  A set of astronomical sources with known
physical fluxes (or more importantly known physical colors) are
established as ``fundamental standards''.  A network of fainter
primary and secondary standard stars, accessible to SNAP and larger
ground-based telescopes, are established with respect to the
fundamental standards.  The SNAP observatory has an observation and
analysis program that takes standard star observations and derives a
transformation between SNAP signals and fluxes in the instrument
passbands.  These programs are described in detail in the following
subsections.

\subsection{Fundamental Standards}

In the ideal situation, target supernovae would be calibrated directly
against a true irradiance standard observed through the same optical
chain.  However, because of the unavailability of convenient
irradiance standards observable through the same optical path as the
supernovae, a suitable, calibrated proxy (or proxies) must be
identified.  Consequently SNAP must use stars as irradiance and color
standards (standard candles).  These selected stars may be established
as standards using calculations of model stellar atmospheres as a
substitute for a NIST-traceable irradiance standard.  A few stars may
be unreddened and easy to model, so that all or part of the standard
candle flux distributions can be replaced by a suitably normalized
model atmosphere calculation.  Traditionally in optical astronomy the
bright, $V=0$ star Vega has been used as the NIST stand-in.  However,
Vega itself is far too bright to be used for most astronomical
programs, so it has been used to calibrate fainter stars e.g. the
Landolt standards, Oke standards etc.  The SNAP calibration program
has identified ``fundamental standard stars'' to be bright, $V \sim 12$,
stars whose spectral energy distributions are independently
determined.  At $V=12-13$, the fundamental stars may be observed with
the SNAP spectrograph.

The current Hubble Space Telescope fundamental flux standards are a
set of three WD stars and could be a good choice for the basis of the
SNAP flux calibration. These stars have pure hydrogen atmospheres and
are too hot for molecular constituents, which means that the physics
of the model calculations are relatively straightforward. The three
models are normalized near 5500 \AA\ using Landolt V band
magnitudes. The HST program (\citep{bohlinetal:2001} also provides
three primary solar analog standards, whose optical flux has been
measured on the WD flux scale and whose IR flux shape is solar. The
stellar model approach is a tested method that yields a precision of
$\sim 1$\% in the UV and optical, while the near infrared relative
fluxes are less well tested.  The systematic flux errors in the WD
model technique have been estimated by the difference between the two
extreme models that still fit the observations. This uncertainty
reaches a maximum in the SNAP wavelength range of 1.8\% at
1.7$\mu$m, meeting the goals of the reference mission.
One possibility for further reducing the uncertainties is to verify
that the WD model $T_{eff}$ and $\log{g}$ values derived from the
Balmer line profiles fit uniformly for the Lyman and Paschen lines, as
well. A second verification in progress with NICMOS is to confirm to
$<1$\% the relative IR fluxes of the models and the solar analogs.

The direct method of determining stellar fluxes is to compare
observations of stars, particularly in the near infrared, with a lamp
source.  As stated in \citet{bohlinetal:2001} ``In order to improve
the accuracy of the standard star IR flux distributions, some standard
stars should be compared to an actual standard IR lamp to define a
proper absolute flux calibration, just as was done for Vega. \ldots
such fundamental data are lacking \ldots'' These authors normalize the
flux scale at one bandpass once the spectral shape has been
determined, while a lamp comparison determines both the absolute level
and the shape of the spectral flux distribution.  The absolute
calibration of Vega
\citep{hayes_latham_75,hayes:85}
exemplifies a direct method of establishing a fundamental calibrator.
The need and approach for such an experiment is under study by the
SNAP calibration team.

\subsection{SNAP Standard Star Network}
In order to minimize the need for on-orbit calibration time, a
ground-based standard star network consisting of the fundamental
standards linked to primary and secondary standards will be
established. This network will provide the preliminary calibration
data for the spacecraft by setting the spectrophotometric
zeropoints. Further, the network will enable parallel and follow-on
observation from the ground to support on-orbit science observations.
The network will establish an all-sky photometric system covering the
spectrum from 3500 to 17000 \AA\ and spanning approximately 6
magnitudes (12 to 18 mag) of brightness.

Primary standards are fainter ($V=15-18$) than the fundamental
standards and will be calibrated with respect to the fundamental
standards through spectrophotometry and photometry.  They will be used
to transfer the calibration to an ``all sky'' grid.  Some will lie in
the SNAP science fields and others around the celestial equator to tie
the northern and southern SNAP fields together and provide standards
distributed over 24 hours.  The primary standards will be tied to the
secondary standards through spectrophotometry and photometry.
Secondary standards will be in the SNAP fields.  They will span a wide
color and magnitude ($V > 18$) range, and hence spectral type.
Uncertainties will be minimized by multiple observations of the
fields.

Once in orbit, we will verify the zeropoints for both the primary and
the secondary standard stars.  The planned repeating observation
cadences will allow typically 140 observations of every star in the
SNAP-N and SNAP-S fields over the course of the mission, leading to
refinement of the brighter secondaries and establishment and
verification of the fainter secondaries.

\subsection{Calibration Transfer}
SNAP is developing a flux-based (rather than magnitude-based) plan for
transferring the fundamental standard star calibration to the primary
and secondary standards, and ultimately to supernovae through a
combination of spectroscopy and photometry.  There are two common
choices of calibration passbands.  Traditionally in observational
astronomy a set of passbands is associated with a specific magnitude
system.  These (virtual) passbands, when used to observe the system's
primary standard stars, reproduce exactly the tabulated magnitudes.
\citet{bessell_filtdef_90} has produced throughput realizations of the
Johnson-Cousins magnitude system.  Another approach is to use
observations to infer the actual throughput of the observers optical
system (including atmosphere).  SNAP can use both approaches in its
calibrations.  Further, as the demand for precise calibration is
fairly stringent, SNAP needs to be able to account for changes in the
instrument performance.  For example, we intend to re-determine the transmission properties of
the photometry filters during the mission by taking advantage of the
multiple observations of stars in the SNAP fields provided by the 4-day
observing cadence.  Internal calibrated,
stable illumination sources may also be utilized.

\section{SNAP Simulation}
\label{sim:sec}
The design of the SNAP reference mission has been guided by
the Fisher-matrix analysis described in \S\ref{asprobe}.
A detailed simulation of the reference mission now provides
validation of the scientific reach of the SNAP mission and
provides a tool for further mission refinement and optimization.
We have developed a detailed parameterized simulator of supernova
cosmology missions.
We describe here its use in projecting
the scientific yield from the SNAP supernova reference mission
described in this paper.  A similar but more complete
pixel-level simulation will be used to test the mission-analysis
software.

For the telescope/camera configuration and observing plan described in
the previous sections, we generate a list of Type Ia supernovae and
their associated host galaxies in the observed field of view during
the course of the survey.  The incident flux at the telescope is
calculated for each supernova, accounting for the underlying
cosmology, magnification from gravitational lensing, and host and
Galactic dust.  The photometric observations of the supernovae from
the SNAP scanning strategy are simulated, resulting in multi-band
light curves for each event.

The rest-frame $B$ and $V$ light curves are fit to SN Ia-class
templates while the light curves in other filters are fit
to a polynomial, in order to measure the magnitude, color, and light-curve
parameters of each supernova.  Triggered spectroscopic observations
are simulated through Fisher-matrix techniques
\citep{bk:03} which predict the quality of the spectroscopic
parameter measurements.  Supernova observables
are related to the intrinsic
peak absolute magnitude using the models described in \citet{linder:hoeflich}.
In this manner, supernova's light-curve and spectral
parameters are then used to determine its distance modulus,
along with host-galaxy extinction ($A_V$ and $R_V$ of the
\citet{card89} dust model) and its individual absolute magnitude.

Given the reference SNAP supernova distribution and expected distance
moduli folded with the expected supernovae from the Nearby Supernova
Factory, we fit the cosmological parameters.  While this can be done
for a general set of parameters, we concentrate on the dark-energy
parameters $w_0$ and $w'$ assuming a flat universe and a fiducial
model with $w_0=-1$ and $w'=0$ (recall we define $w'\equiv dw/d\ln
a|_{z=1}
\approx w_a/2$). The standard systematic-error model is adopted.
Marginalizing over all other parameters, we get uncertainties in $w_0$
of 0.07 and $w'$ of $\sim0.31$ for the prior 0.03 uncertainty in
$\Omega_M$.  If we take no prior in $\Omega_M$ but
assume cosmological constraints expected from Planck, the $w'$
uncertainty improves to $\sim 0.16$.  These results are somewhat
better than those determined in the Fisher analysis shown in
Table~\ref{sci_errors.tab} where all supernovae are assumed to have
identical distance-modulus uncertainty.  By setting exposure times
based on light-curve requirements for the highest redshift supernovae,
the lower redshift events have better $S/N$ with the fixed filter
design.

\begin{figure}[h!]
\plotone{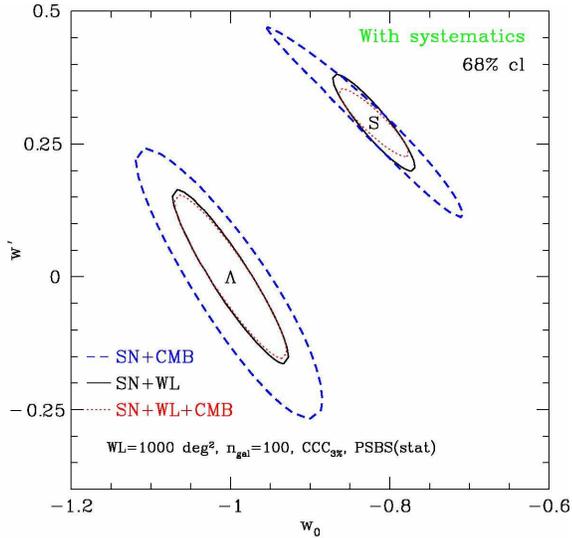}
\caption{The 68\% confidence region in the dark-energy parameters
for fiducial cosmological constant (labeled with $\Lambda$) and
supergravity (labeled with S) universes.  The blue dashed curves show
the results of the supernova simulation with a Planck prior and
an assumed flat universe. Using
Fisher analysis, we find significant improvement is obtained using the
SNAP supernova and 1000 square-degree weak lensing surveys (black
curves).  Additional CMB data (dotted red curves) provides additional
constraints for the supergravity universe, but not for the fiducial
cosmological model. We define $w'\equiv dw/d\ln a|_{z=1}$.
\label{w0w':fig}}
\end{figure}

To explore how SNAP will be able to distinguish between different
dark-energy models, we simulate constraints on $w_0$ and
$w'$ choosing
as fiducial models instantiations of two postulated theories that
account for an accelerating universe: a cosmological constant and a
supergravity-inspired model.  Using Monte Carlo we generate 68\% confidence
contours for SNAP supernovae, in
conjunction with the data anticipated from the Planck mission,
and SNAP supernovae and weak lensing.
(Figure~\ref{w0w':fig}).  We marginalize over the absolute magnitude
and $\Omega_M$ with no prior imposed on either.

Even with just SNAP supernovae, the chosen models are distinct, with
the fiducial values lying well outside the 68\% contours of the other
model.  Inclusion of SNAP weak-lensing power spectrum, bispectrum, and
cross-correlation cosmography information provides improved
sensitivity.  SNAP and complementary data will provide important
guidance to the true nature of dark energy.

We have done the most detailed simulation of a weak lensing survey to
date by identifying instrumental aspects of SNAP that will affect the
PSF, applying this PSF to simulated images, and calculating the
lensing efficiency and subsequent constraints on cosmological
parameters from the SNAP weak lensing survey.  The results of these
studies are in
\citet{rhodesetal:2003, masseyetal:2003,refregieretal:2003}. 
The cosmological reach of the SNAP weak-lensing program presented in
this paper is based on these simulations.  We are continuing to
develop the image simulations for more accurate results and to better
identify instrumental factors that might be tweaked for gains in
sensitivity to weak lensing.  We also are exploring the impact of new
analysis techniques, e.g.\ bispectrum measurements of the mass
distribution and cross-correlation cosmography, for measuring
dark-energy parameters \citep{bernsteinjain:2003,jaintaylor:2003,
takada:2003:mnras}.

\section{Ancillary Science with the SNAP Survey Fields}
\label{ancsci:sec}
The SNAP surveys will have an unprecedented combination of depth,
solid-angle, angular resolution, temporal sampling, and wavelength
coverage from the optical to the NIR.  The wealth of potential
science from  deep and wide-field surveys has already been demonstrated
in such surveys as the Hubble Deep Fields \citep{williams_96,williams:2000}
and the SDSS \citep{sdss:2003}.  In \S\ref{otherde:sec} we summarized
how the SNAP data set will contribute to our understanding of dark energy.
We here explore the properties of
the surveys and their potential scientific yield in other areas.
\subsection{SNAP Survey Depth}
\label{depth:sec}

Based on the telescope and camera characteristics, we quantify the
expected depth, solid-angle, and time resolution of the SNAP supernova and
weak lensing surveys.  The telescope and camera properties of SNAP
have been modeled and incorporated into an advanced exposure-time
calculator (ETC) \citep{bernstein:etc}.  Besides having all the usual
capabilities of a standard ETC, our ETC includes unique handling of
the pixel response function, undersampling, dithering, and
probabilistic cosmic-ray rejection.  As mentioned in
\S\ref{imager:sec}, SNAP will use dithering to obtain spatial
resolution below the pixel scales.

The high cosmic-ray flux in a space environment produces a non-trivial
reduction of effective exposure times; pixels from a single exposure
that are contaminated by a cosmic ray are recognized
through median and other filtering and dropped in the dithered
reconstruction.  The expected contamination
probability of a point source is 0.0002 per second.
Short individual exposure (plus readout) times  limit the
contamination: four 300 second exposures give better than a 99\%
probability that there will be no cosmic-ray contamination at a given
position for at least one of the dithers that make up a pointing.

The magnitude depths for individual scans and co-added images of the
SNAP supernova fields are calculated for each filter.
Table~\ref{depth:tab} shows the limiting $S/N=5$ AB magnitude for each
filter in the surveys, for point sources that are not contaminated
with cosmic rays. As with the HST, the limiting magnitude for any
given point is probabilistic, due to the random occurrence of cosmic
rays.  The probability for a given level of cosmic contamination (and
approximate degradation in signal-to-noise) can be determined using
the binomial distribution.  In a single scan of the deep survey 77\%
of point sources will have no cosmic contamination in any of the four
dithered images while 98\% will have one or fewer cosmic-contaminated
image, only slightly reducing signal-to-noise.

\begin{deluxetable}{cccccccc}
\tablewidth{0pt}
\tablecaption{The SNAP
AB magnitude survey depth for a point source $S/N=5$.\tablenotemark{a}
\label{depth:tab}}
\tablehead{
Filter & $\lambda_{eff}$(\AA) & $\Delta \lambda$(\AA)& \multicolumn{2}{c}{Deep Survey (AB mag)} & Wide-field Survey & Panoramic\\ \cline{4-5}
& & &Per Scan & Co-added Scans& (AB mag)}
\startdata
      1&        4400&        1000&   27.9&   30.6&   28.3&   26.7\\
      2&        5060&        1150&   27.9&   30.6&   28.3&   26.8\\
      3&        5819&        1322&   27.8&   30.5&   28.2&   26.8\\
      4&        6691&        1520&   27.8&   30.4&   28.1&   26.8\\
      5&        7695&        1749&   27.8&   30.4&   28.1&   26.8\\
      6&        8849&        2011&   27.7&   30.3&   28.0&   26.7\\
      7&       10177&        2313&   27.5&   30.1&   27.8&   26.6\\
      8&       11704&        2660&   27.5&   30.1&   27.8&   26.6\\
      9&       13459&        3059&   27.4&   30.0&   27.7&   26.5\\ 
\enddata
\tablenotetext{a}{Random cosmic-ray hits make the $S/N$ for a given position probabilistic; see text.
The choice of filter set is currently subject to optimization studies;
the filters and depths presented here are meant to be illustrative.}
\end{deluxetable}

The SNAP observing strategy provides remarkably even depth over
the range of filters.  For a given filter,
{\it individual scans}
 of the deep and wide-field surveys are only $\sim 0.75$
magnitudes shallower than the Hubble Deep Fields (HDFs) while the deep fields
co-added over time are
$\sim 1.5$ magnitudes deeper than the HDFs \citep{williams_96}
and up to 0.5 magnitudes deeper than the UDF depending on the filter.
SNAP has the additional advantage of
having nine filters observing to this depth, compared
to the four filters of the HDFs, and 9000 times the area in the deep
survey;
when these data from all filters are combined,
the limiting magnitude increases by another 0.6 magnitudes.
The wide-field survey has $\sim 600,000$ times the area of the 
HDFs.

SNAP fields will contain many faint diffuse galaxies whose detection
is important for the weak-lensing studies, and for other potential science
projects.
The limiting magnitudes for Gaussian-aperture photometry of an exponential-disk
galaxy with FWHM=0.12'' are shown in Table~\ref{depth_diff:tab}.

\begin{deluxetable}{cccccccc}
\tablewidth{0pt}
\tablecaption{The SNAP AB magnitude survey depth for an exponential-disk galaxy
having FWHM=0.12'' for $S/N=10$.\tablenotemark{a}
\label{depth_diff:tab}}
\tablehead{
Filter & $\lambda_{eff}$(\AA) & $\Delta \lambda$(\AA)& \multicolumn{2}{c}{Deep Survey (AB mag)} & Wide-field Survey & Panoramic\\ \cline{4-5}
& & &Per Scan & Co-added Scans& (AB mag)}
\startdata
      1&        4400&        1000&   26.4&   29.0&   26.8&   25.2\\
      2&        5060&        1150&   26.3&   28.9&   26.7&   25.3\\
      3&        5819&        1322&   26.3&   28.9&   26.6&   25.3\\
      4&        6691&        1520&   26.2&   28.8&   26.6&   25.3\\
      5&        7695&        1749&   26.3&   28.9&   26.6&   25.3\\
      6&        8849&        2011&   26.2&   28.8&   26.5&   25.3\\
      7&       10177&        2313&   26.2&   28.8&   26.5&   25.3\\
      8&       11704&        2660&   26.2&   28.8&   26.5&   25.3\\
      9&       13459&        3059&   26.1&   28.7&   26.5&   25.3
\enddata
\tablenotetext{a}{Random cosmic-ray hits make the $S/N$ for a given position probabilistic; see text.  The choice of filter set is currently subject to optimization studies;
the filters and depths presented here are meant to be illustrative.}
\end{deluxetable}

\subsection{Ancillary Science}
\label{science:sec}
In this section we give a brief outline of additional science topics
likely to benefit from SNAP survey data. This list is incomplete, and
includes only a few of the most obvious ancillary science projects. 

The SDSS \citep{York:2000},
Hubble Deep Fields (HDFs) \citep{williams_96,williams:2000},
and Ultra Deep Field\footnote{http://www.stsci.edu/hst/udf} 
demonstrate the vast range of science enabled by wide, deep, and 
precise multi-band surveys. SNAP will carry this tradition forward
as the first wide field survey in space.

SNAP deep-survey fields will cover more than 9000 times the area of
the HDFs, go even deeper, include 9 color observations, and provide
significant time domain information.  The SNAP wide-survey field is
comparable in size to the Sloan Southern
Survey\footnote{http://www.sdss.org} and CFHT Legacy
Survey\footnote{http://www.cfht.hawaii.edu/CFHLS} but includes
diffraction limited imaging, and probes many magnitudes fainter in
more filters.  A panoramic survey would cover almost a quarter of the
full sky at slightly shallower depth.  The combination of depth,
temporal coverage, wavelength range, diffraction-limited seeing, and
wide field make SNAP imaging surveys uniquely powerful in the study of
a wide range of objects and phenomena, which we outline below.

\begin{itemize}
\item Galaxies --- Within the 15 square degree deep
survey area, SNAP will make accurate ($\Delta_z < 0.03$)
photometric redshift estimates
for at least $10^{7}$ galaxies from redshift 0 to 3.5, spanning
more than 90\% of the age of the Universe.  Statistical studies are
possible with such a large sample, e.g.\ the determination of the
galaxy luminosity function and color distributions as a function of
redshift.  Photometric redshifts can be estimated from the 4000~\AA\
break for galaxies out to about $z = 3.2$.  For galaxies at still
higher redshift, the simplest indicator is the Lyman break. For SNAP,
the Lyman break enters the optical imager around redshift 3. In
principle it can be followed using SNAP data beyond redshift 10,
allowing identification of extremely high-redshift galaxies.  The
magnitude
depth also allows discovery of low-surface-brightness and very
high-redshift galaxies.  High-resolution images will provide a view of
the internal structure of galaxies and their interactions
with each other.  This data set, which will include morphological
information for every object, will provide a unique opportunity to
study the evolution of galaxies.  The flood of galaxy evolution papers
based on the Hubble Deep Fields points the way to what will be possible
with the SNAP imaging data set.

\item Galaxy structure formation --- The wide-field surveys from SNAP offer an
excellent opportunity to study galaxy clustering and its evolution 
at high redshifts ($\approx 0.3$ -- 3). This is a regime which is both
especially interesting and rather poorly understood. There are some 
measurements of galaxy clustering at high redshift \citep{Steidel:1996},
but these are severely limited by sample variance \citep{Somerville:2003}.
SNAP, increasing the size of high redshift surveys by several orders of
magnitude, will be substantially less sensitive to cosmic variance.
Photometric redshifts greatly enhance this work, allowing detailed study 
of the evolution of galaxy clustering.

\item Galaxy clusters ---
Galaxy clusters, the most massive bound objects in the Universe,
provide important probes of our understanding of structure
formation \citep{Evrard:1989}. 
Constraining their formation and evolution is an important
observational goal for the coming decade. Recent advances have
overcome earlier limitations of optically selected cluster samples,
essentially by using photometric redshift information to eliminate
projection effects \citep{Gladders:2000,Bahcall:2003}. 
Hubble Volume $\Lambda$CDM simulations \citep{Evrard:2002} predict
that the SNAP surveys will provide detailed information
on tens of thousands of galaxy clusters with masses above $5\times10^{13}$
M$_\odot$. With NIR sensitivity, SNAP will allow sensitive detection of 
galaxy clusters all the way back to their earliest appearance around 
$z = 2$. The overall number of galaxy clusters is exponentially
sensitive to the amplitude of initial density fluctuations.

\item Quasars --- Quasars are identified in multi-color imaging
surveys by their non-stellar colors. This method has been shown by the
Sloan Digital Sky Survey to be extremely effective at identifying
quasars to redshift 6 and beyond. SDSS quasar discovery is limited to
redshift 6.4 by the CCD sensitivity cutoff at 1.0 $\mu$m \citep{pen02}.
The most distant SDSS quasar, at redshift 6.42 \citep{Fan:2003}, 
has a $z$-band magnitude of 20. By probing to
wavelengths 1.7 times greater, and to depths 9 magnitudes fainter,
SNAP will be able to detect quasars to redshift 10, and to probe
the quasar luminosity function to 100 times fainter than the brightest
quasars.   SNAP's ability to identify diffuse objects associated with
quasars may present interesting opportunities for the study of galaxy
formation.

\item Gamma-ray burst afterglows --- Current evidence suggests
that gamma-ray bursts are associated with the collapse of massive
stars which live short lives and die where they are born. As a result,
GRB's may trace the cosmic star formation rate.  If so, there should be
GRB's essentially coincident with the epoch of formation of the first
stars. The most distant GRB known occurred at redshift of 
4.5 \citep{Anderson:2000}. SNAP will be able to identify GRB afterglows, 
and the orphan afterglows
predicted by some models of beaming in GRB's to $z = 10$. Such orphan
afterglows may even be detected as backgrounds to the SNe search.

\item Reionization history ---  The Universe became
neutral at the time of recombination, around $z = 1000$, and the thermal
radiation from that epoch travels to us undisturbed as the cosmic
microwave background radiation. The lack of a Gunn-Peterson effect in
the spectra of most quasars demonstrates that the Universe was
reionized at some time between $z = 1000$ and $z = 6$. The source of the
ionizing radiation is the subject of substantial speculation.
The recent discovery of an apparent Gunn-Peterson trough in the spectra
of several $z > 6$ SDSS quasars may provide the first glimpse of the
epoch of reionization \citep{White:2003}.  
By identifying many quasars and galaxies to
$z=10$, SNAP will set the stage for mapping the epoch of reionization in
unprecedented detail. In combination with ground based and JWST
spectroscopy, it will enable measurements of the proximity effect and
studies of the spatial structure of reionization.

\item Transients/Variables --- The discovery and observation
of SNe Ia are the primary goals of SNAP, but transient ``backgrounds''
are interesting in their own right: quasars, active-galactic-nuclei,
gamma-ray-burst
optical counterparts, supernovae of other types, variable stars,
and eclipsing binaries.  Of particular interest to cosmology are 
time-delay measurements of the
large expected number of strongly lensed variables \citep{Blandford:2001}.
Gravitational microlensing studies of stars and quasars as probes of 
compact dark matter are also possible.

\item Stars --- Faint limiting magnitudes and excellent star-galaxy
separation will yield faint dwarf and halo stars.  Proper motion
can be detected with high-resolution and a long time reference.
SNAP's accurate colors will yield excellent photometric parallaxes
to all stars in the field.
The geometry and substructure of the Galactic halo and disk in the direction
of the SNAP fields can be mapped.
Of particular interest would be a
census of low-mass L and T stars and brown dwarfs throughout the Milky
Way disk \citep{leg00}.
\item Solar-system objects ---  The peculiar motion and parallax
in the time-series
data will facilitate the identification of local objects such as
asteroids and Kuiper-belt objects.  SNAP time series data will provide
an excellent probe of faint, red objects in the Kuiper belt and
beyond.  A 2-3 month dedicated ecliptic survey with the SNAP survey
would detect 10--50,000 Kuiper belt objects down to the size of the
Comet Halley's nucleus.

\item Weak gravitational lensing ---
The importance of weak gravitational lensing for dark energy
is described in \S\ref{lensing:sec}.
Weak gravitational lensing is also a powerful tool in mapping the
distribution of mass in the Universe.  Maps reconstructed from the
observed shear field are sensitive to any mass along a given line of
sight, regardless of its nature or state allowing dark matter mapping
and the unbiased detection of galaxy clusters and their mass. The high
surface density of resolved background galaxies enables the
construction of both two- and three-dimensional maps of the foreground
dark matter distribution.  In addition to providing further
constraints upon cosmological parameters, these maps trace the growth
of structure in the Universe and can be use to study the bias between
mass and light through comparison with the galaxy distribution.

In recent years many groups have
measured the shape distortions of field galaxies due to weak lensing
and used these shear measurements to set constraints on cosmological
parameters. In a number of weak-lensing surveys using ground and
space-based data the second moment of the shear field has provided a
constraint on the matter power spectrum, in particular on the quantity
$\Omega_M \sigma_8^{0.5}$, that is comparable to the constraints set
by more traditional methods such as the abundance of X-ray clusters.
Current weak-lensing surveys seeking to measure higher order moments
of the shear field and the magnification of background galaxies
promise to provide an independent measure of $\Omega_M$, thus breaking
the degeneracy between $\Omega_M$ and $\sigma_8$, where $\sigma_8$ is
the amplitude of mass fluctuations on the scale $8h^{-1}$ Mpc.

Precise photometric redshifts enable the making of 2-dimensional mass
maps in multiple redshift slices, and even 3-dimensional mass
reconstructions. This redshift tomography is much less effective from
the ground because of the lower surface density of resolved galaxies,
and cannot be extended as far.  Space-based imaging is needed to
resolve sufficiently many source galaxies at $z>1$ and follow the
growth of structure at those redshifts. Furthermore, high-resolution
images coupled with photometric redshifts permit the use of a recently
formulated approach to 3-D mapping \citep{taylor:2001} which recreates
the full 3-D mass distribution by simultaneously considering both the
shear estimator and the photometric redshift of a galaxy. Applied to
the SNAP deep survey, this technique will detect mass overdensities
with a $1\sigma$ sensitivity of $10^{13}M_\sun$ (1/5 the mass of the
Virgo cluster) at $z=0.25$.

\item Strong gravitational lensing --- The high spatial resolution of
SNAP NIR observations will enable the discovery of a large number of
new strong lenses \citep{kuhlenetal:2003}. The statistics from such a
large, uniformly selected sample would be ideal to impose constraints
on the density profiles of galaxies and clusters, as well as to
constrain cosmological parameters beyond those of dark energy.  The
NIR observations, which are much less sensitive to dust extinction
within the lens galaxy, are especially important in regard to strong
lensing observations.
\citet{goobaretal:2002} find that assuming a
flat universe, the statistical uncertainty on the mass density is
$\sigma^{\rm stat}_{\Omega_M} \ls 0.05$, although uncertainty of the
lens modeling are likely to dominate.

\end{itemize}

The SDSS demonstrates clearly the way in which
a wide-field survey can produce scientific yield well beyond
its primary design goals.
Individual objects found on SDSS images are routinely observed
spectroscopically at the largest telescopes in the world,
fulfilling the
historical trend of wide-field small-aperture telescope imaging feeding
targets
for large-aperture telescope spectroscopy.
The SNAP surveys will provide a similar opportunity in working
with JWST and the next generation of ground-based
extremely large telescopes.

\section{Conclusion}
\label{summary:sec}

The discoveries of recent years make this a fascinating new era of
empirical cosmology, in which we are developing new
technologies and instrumentation
capable of addressing fundamental questions.
The extraordinary challenge of the precision and
accuracy necessary to explore the properties of dark
energy requires a new stringent data set in order to improve
statistical uncertainties and control
systematic uncertainties; large numbers of supernovae over an
extended redshift range with light curves and spectra measured to high
signal-to-noise and deep, wide fields full of weak-lensed galaxies
with high-resolution PSFs with unprecedented stability.

Based on these data requirements, we have developed a
reference mission called the Supernova /
Acceleration Probe,
consisting of a detailed observing
program carried out with an optimized wide-field telescope and
attendant imaging and spectrograph instruments.
The optical, mechanical, and thermal studies
and analyses conducted to date indicate that the reference model is
manufacturable and testable using proven techniques. The imager will
provide PSF-controlled wide-field images
of galaxies with efficient multiplexed supernova discovery and light-curve
building while the spectrograph satisfies the performance
requirements for measuring key spectral features of the
faintest supernovae.

The reference model presented in this paper addresses the many
requirements necessary for a high-precision supernova experiment,
a PSF-controlled weak-lensing experiment, and makes possible a range
of further complementary dark-energy studies.  We
use numerical simulations and trade studies to refine the mission concept
and quantify the expected science yield of SNAP.  This paper will
be followed up with further results of these
studies and the progress in the optimization of SNAP.

Although SNAP is designed as a self-contained experiment, its surveys
can be used in conjunction with complementary ground-based
and JWST-based studies that can further strengthen the reference SNAP
mission.

SNAP presents a unique opportunity to probe the dark energy and
advance our understanding of the Universe.  Moreover, the principal survey
missions of SNAP will produce a cornucopia of observations
capable of revolutionizing other areas of astrophysics and cosmology.
After completion of these primary missions, a guest survey
program is envisioned filling the remaining satellite lifetime to
allow the full potential of SNAP to be realized.

\section*{Acknowledgments}
We thank Masahiro Takada for making detailed results of his work
available.
This work was supported by the Director, Office of Science, of the
U.S.   Department of Energy under Contract No. DE-AC03-76SF00098.



\end{document}